\documentclass[a4paper,10pt]{article}
\usepackage{makeidx}
\usepackage[english]{babel}
\usepackage[latin1]{inputenc}
\usepackage{amsfonts}
\usepackage{hyperref}
\usepackage{bm}
\usepackage{amssymb}
\usepackage{latexsym}
\usepackage{amsmath}
\usepackage{amscd}
\usepackage{amsthm}
\usepackage{mathrsfs}
\usepackage{amsmath,amstext}
\usepackage{graphicx}
\usepackage{color}
\usepackage{fancyhdr}
\usepackage{epsfig}
\usepackage{alltt}

\theoremstyle{plain}
\newtheorem{theorem}{Theorem}[section]
\newtheorem{conjecture}{Conjecture}[section]
\newtheorem{prop}[theorem]{Proposition}
\newtheorem{lemme}[theorem]{Lemma}
\newtheorem{corol}[theorem]{Corollary}

\newtheorem*{Thm*}{Theorem}

\newtheorem*{Pte}{Proposition}

\theoremstyle{definition} 
\theoremstyle{definition} 
\theoremstyle{definition} 
\newcommand{\be} {\begin{equation}}
\newcommand{\ee} {\end{equation}}
\newcommand{\bea} {\begin{eqnarray}}
\newcommand{\eea} {\end{eqnarray}}
\newcommand{\Bea} {\begin{eqnarray*}}
\newcommand{\Eea} {\end{eqnarray*}}

\title{Branching Feller diffusion for cell division with parasite infection}

\author{Vincent Bansaye\thanks{CMAP, Ecole Polytechnique,  Route de Saclay,
91128 Palaiseau Cedex, FRANCE. vincent.bansaye@polytechnique.edu}, Viet Chi Tran\thanks{Laboratoire Paul Painlevé, Université des Sciences et Technologies de Lille.
59 655 Villeneuve d'Ascq Cedex,  chi.tran@math.univ-lille1.fr}}

\date{\today}

\numberwithin{equation}{section}

\def\t{\textrm}
\def\P{\mathbb{P}}
\newcommand{\E}{\mathbb{E}}
\newcommand{\D}{\mathbb{D}}
\newcommand{\Co}{\mathcal{C}}

\newcommand{\iid}{\mbox{i.i.d.} }
\newcommand{\lhs}{\mbox{l.h.s.} }

\newcommand{\as}{\mbox{a.s.} }
\newcommand{\rv}{\mbox{r.v.} }

\newcommand{\N}{\mathbb{N}}

\newcommand{\R}{\mathbb{R}}

\def\eg{\textit{e.g.} }

\def\N{\mathbb{N}}

\def\P{\mathbb{P}}
\def\R{\mathbb{R}}
\def\E{\mathbb{E}}

\def\t{\textrm}
\def\w{\widetilde}

\def\d{\emph{d}}

\def\ind{{\mathchoice {\rm 1\mskip-4mu l} {\rm 1\mskip-4mu l}
{\rm 1\mskip-4.5mu l} {\rm 1\mskip-5mu l}}}

\theoremstyle{remark}

\begin{document}

\maketitle

\begin{abstract}We describe the evolution of the quantity of parasites in a population of cells which divide in continuous-time.
The quantity of parasites in a cell follows a Feller diffusion, which is splitted randomly between the two daughter cells
 when a division occurs.  The cell division rate may depend on the quantity of parasites inside the cell and we are interested
in the cases of constant or monotone division rate.  We first determine the asymptotic behavior of
the quantity of parasites in a cell line, which follows a Feller diffusion with multiplicative jumps. We then
 consider the evolution of the infection of the cell population and give  criteria to determine whether the proportion of infected cells goes to zero (recovery) or if a positive proportion of cells becomes largely infected (proliferation of parasites inside the cells).
\end{abstract}


\noindent \textbf{A.M.S. classification: }60J80, 60J85, 60J60, 60J75, 92D25.\\

\section{Introduction and main results}

 We consider a continuous time model for dividing cells infected by parasites. We assume that parasites proliferate in the cells and that their lifelengths are much less than the one of the cell. Informally, the quantity of parasites $(X_t : t\geq 0)$ in a cell evolves as a Feller diffusion \cite{fellerlivre}. The cells divide in continuous time with a rate $r(x)$ which may depend on the quantity of parasites $x$ that they contain. When a cell divides, a random fraction $\Theta$ of the parasites goes in the first daughter cell and a fraction $(1-\Theta)$ in the second one. More generally, instead of parasite infection,  one could be interested in some biological content which grows in the cells and is shared randomly when the cells divide (for example  proteins, nutriments, energy or extrachromosomal rDNA circles in yeast \cite{sinclairguarente}). \\

 Let us give some details about the biological motivations.
 For the sharing of parasites, we are inspired by experiments conducted in Tamara's Laboratory where bacteria E-Coli have been infected with bacteriophage lysogens \cite{tad}. They show that a very infected cell often gives birth to a very infected and a lowly infected daughter cells. Thus we are interested in taking into account unequal sharing and we do not make restrictive assumptions about $\Theta$. The model we consider
here is a continuous  version of Kimmel's multilevel model for plasmids  \cite{kimmel}. In the latter model,
the cells divide in continuous time at a constant rate and the number of parasites is a discrete quantity which is fixed at the birth of the cell: the parasites reproduce 'only when the cells divides'. Moreover the sharing of parasites is symmetric. In  \cite{bansaye}, a discrete time model where the sharing of the parasites may be asymmetric is considered. We refer to \cite{atarbre,benj} for general discrete time models with independent and identically distributed values   for the offspring and to \cite{guyon, delmasmarsalle, bercugegoutpetitsaporta} for asymmetric models motivated by cellular aging. Here both the quantity of parasites and the time are continuous.\\
In addition to unequal repartition of parasites in the two daughter cells, we are also interested in letting the division rate of the cell $r(x)$ depend on the quantity $x$ of parasites in the cell. On the one hand the cell's division rate may decrease when the quantity of parasites increases  since the cell's stamina is affected. For instance, in \cite{Bonds2006, Hurd2001, Okeefe2002} examples where host fecundity is reduced by the presence of pathogens are presented. On the other hand, increasing rates may be found in order to fight against their proliferation. Furthermore, this case is  relevant when there is a symbiosis between the host and its parasites: parasites then make the rate of division of the cell increase. In this direction, we refer to \cite{bull1991, bull1992}. Note that increasing functions $r$ are also natural if  we replace parasites by nutriments or more generally 'energy for the cell division'. Finally, we also consider the case where the division rate does not depend on the quantity of parasites ($r$ constant) since some parasites do not affect the evolution of their host. Moreover this is mathematically natural to consider this simpler case first, which will be useful for non constant cases. \\
The process we study is a Markov process on Galton Watson trees and \cite{bansayedelmasmarsalletran} give asymptotic results under an ergodicity assumption, which is not fulfilled here. Mathematical approaches for the case of non constant division rates $r(x)$ have been considered by \cite{hardyharris3} for continuous time branching diffusions, when the offspring appear at the position of their mother. Their method, which relies on Girsanov transformations, does not hold when nonlocal branching is considered. In this vein, we also refer to \cite{Engl, Engll}. \\
  As we will observe, even when the division rate is constant, our model behaves differently from the discrete time model with 'constant division' of \cite{bansaye} and the recovery criteria are different.\\
 We refer to \cite{dawson,dawsongorostizali} for the nonlocal branching superprocess counterpart. In the framework of cell division, \cite{evanssteinsaltz} considers a superprocess approach for cell damages, which corresponds to the high density limit of small and short living cells. Notice however that asymmetry and nonlocal branching are lost in the continuous limit. Moreover, in view of statistical applications \cite{guyon, delmasmarsalle}, our purpose is to stick to a discrete genealogical cell tree. In the experiments, the number of cells can not always be considered as large.  \\

Let us give now a  qualitative description of our model, which is rigorously defined in  the next section. The quantity of parasites $(X_t : t\geq 0)$  follows  a Feller diffusion (see e.g.  \cite{lamperti,bingham},  \cite{revuzyor} (Chapter IX) and  \cite{ikedawatanabe} (Chapter IV))
\begin{align}
\d X_t= & g X_t dt+  \sqrt{2 \sigma^2 X_t}\d B_t\label{Fellerdrifted}
\end{align}started at $x_0>0$. In this work, we study the super-critical case ($g>0$) with positive variance $(\sigma>0$). Then
 the parasites survive with  probability $1-\exp(-gx_0/\sigma^2)\in (0,1)$. Upon survival their quantity goes to infinity. This model corresponds to the high density limit of a birth and death process for parasites with short lives and we give details of the convergence of the discrete model to the continuous model studied here in the next section.\\
The population of cells remains discrete and each cell divides in continuous time with a rate $r(x)$ which will be here monotone, measurable and bounded on compact intervals to avoid degenerated situations. Let us denote by $V_t$ the set of living cells at time $t$, by $N_t=\# V_t$ the cell population size at time $t$ and
by $X^i_t$ the quantity of parasites in the cell $i\in V_t$ at time $t$. Thus:
$$\sum_{i\in V_t}X^i_t=X_t.$$
\par When a cell containing a quantity $x$ of parasites divides,  the quantity of parasites of the two daughter cells received by inheritance are  respectively $\Theta x \quad \t{and} \quad (1-\Theta)x$, where $\Theta$ is a random variable in $[0,1]$ such that $\P(\Theta=0)=\P(\Theta=1)=0$.  
 Let us  note that this model can been seen as a random fragmentation where the mass of the fragment follows a Feller diffusion. A fragment with mass $x$ splits with rate $r(x)$ in two fragments whose mass are respectively equal to $\Theta x$ and $(1-\Theta)x$. Let us notice similarities with \cite{brennandurrett1,brennandurrett2} for splitting intervals of size $x$ at rate $r(x)=x^{\alpha} (\alpha>0)$, which amount to considering constant quantity of parasites within cells.\\


\par We aim at determining  how the infection  evolves in the cell population. We stress that the results we give in this paper
do not depend on the initial quantity of parasites $x_0>0$. 

First, in Section \ref{critFm}, we  determine the asymptotic behavior of the quantity of parasites in a cell line $(Y_t:t\geq 0)$. This  amounts to following the infection in a cell until it divides and then choose at random one of the two daughter cells. This process is
a Feller diffusion $X$ with multiplicative jumps $\Theta$ occurring at rate $r(.)$ and we prove the following extinction criterion.

\begin{Pte}\label{propextinction}
(i) In the case where $r(.)=r$ is constant,
\begin{itemize}
\item[$\star$] If $g\leq \E(\log(1/\Theta))r$, then almost surely (a.s.) there exists $t\geq 0$ such that $Y_t=0$.
\item[$\star$] Otherwise $\P(\lim_{t\rightarrow +\infty} Y_t=+\infty)>0.$
\end{itemize}
(ii) In the case where $r$ is an increasing function,
\begin{itemize}
\item[$\star$] If there exists $x_0$ such that  $g\leq \E(\log(1/\Theta))r(x_0)$, then a.s. there exists $t>0$ such that $Y_t=0.$
\item[$\star$] If $g>\E(\log(1/\Theta))\sup_{x\in\R_+}r(x)$, then $\P(\lim_{t\rightarrow \infty} Y_t=\infty)>0.$
\end{itemize}
(iii) In the case where $r$ is a decreasing function,
\begin{itemize}
\item[$\star$] If $g\leq\E(\log(1/\Theta)) \inf_{x\in \R_+}r(x)$, then a.s. there exists $t>0$ such that $Y_t=0.$
\item[$\star$] Otherwise, $\P(\lim_{t\rightarrow \infty} Y_t=\infty)>0.$
\end{itemize}
\end{Pte}
 We say that the organism \emph{recovers} when the proportion of infected cells goes to zero as time goes to infinity.
But contrarily to the discrete case \cite{bansaye, guyon}, the extinction criteria stated above
do not provide directly recovery criteria for the organism. When the division rate is constant, we prove in Section  \ref{section3} the following criterion for a.s. recovery:
\begin{Thm*}\label{threcovery} We assume here that $r(x)=r$ for every $x\geq 0$. \\
(i) If $g\leq  2\E(\log(1/\Theta))r$, then the organism recovers a.s. in the sense that:
\begin{equation}
\lim_{t\rightarrow +\infty}\frac{\# \{ i \in V_t : X^i_t=0\}}{N_t} = 1
\quad \t{a.s.}
\end{equation}
(ii) If $g>  2\E(\log(1/\Theta))r$ then the parasites proliferate exponentially inside  the cells as soon as the parasites do not become
extincted in the sense that
\begin{equation}
\big\{\limsup_{t\rightarrow \infty} \frac{\# \{ i \in V_t : X^i_t\geq e^{\kappa t}\}}{N_t} > 0\big\}=\{\forall t\geq 0, \ X_t
>0\}\quad \t{a.s.}\label{majoprobasurvie}\end{equation}
for every
 $0<\kappa<g-  2\E(\log(1/\Theta))r$.
\end{Thm*}
Thus, when the division rate is constant, either the organism recovers a.s. or a positive proportion of cells becomes arbitrarily largely infected as soon as the total parasite population survives. Let us make two observations about this recovery criterion. First, thanks to unequal sharing of parasites (i.e. $\E(\log(1/\Theta))\gg 1$), the organism may recover a.s. although the parasite growth rate is large ($g \gg 1$) and the cells' division rate is low ($r\ll 1$). Second, we can  remark  the factor \emph{$2$} in this criterion, which is inherited from a bias phenomenon that is typical to continuous time branching process and stated in \cite{bansayedelmasmarsalletran, chauvinrouaultwakolbinger, hardyharris3}. Heuristically, a typical cell at a large time has divided at rate $2r$ along its ancestral lineage. This allows the following surprising situation: the organism may recover a.s. although the quantity of parasites in a cell line  goes to infinity with positive probability. \\


\par The situation is very different if  $x\mapsto r(x)$ varies. In this case, the genealogical tree of the cell population depends on the quantity of parasites that are evolving in the cells and we are no more in the frame of Markov processes indexed by Galton-Watson trees such as
 \cite{bansayedelmasmarsalletran}. \\
 If the division rate decreases when the quantity of parasites increases, then the situation is quite intuitive since the cells which become too infected will keep getting more and more infected whereas  their proportion in the cell population tends to 0. Using the infection in a cell line $(Y_t : t\geq 0)$, we give a sufficient condition for a.s. recovery (see Section \ref{section_decreasing}). \\
The case of an increasing division rate is more interesting and difficult: increase in the cell division rate may prevent the parasites from proliferating in the cells. There is no longer either recovery or proliferation of parasites and an intermediate regime appears (see the first example in Section \ref{section62}). This leads us to introduce \emph{moderate infection}, which roughly speaking means that the distribution of the quantity of parasites in a positive fraction of  the cells alive at time $t$ remains positive and bounded. 
The criterion for this a.s. moderated infection is conjectured in Section \ref{secconj} and illustrated with simulations in Section \ref{sectionsimu}.


The paper is organized as follows.  In Section \ref{section2}, we give more formal definitions based on measured valued processes and state the convergence of the discrete model to the continuous model. In Section \ref{critFm}, we give the asymptotic behavior for  Feller diffusions with multiplicative jumps in the cases of constant or monotone division rates. This gives the asymptotic behavior of the infection in a cell line. We derive
in Section \ref{section3} the recovery criteria for constant division rate rates $r$, while non-constant cases are handled in Section \ref{section4}.

\section{Preliminaries}\label{section2}
\subsection{Measure-valued description of the infected cell population} To describe the tree of the cell population and label its nodes, we use the  Ulam-Harris-Neveu notation (\eg \cite{dawson,legall}). The mother cell is labeled by $\emptyset$ and when the cell of label $i$ divides, the two daughters are labeled by $i0$ and $i1$. We introduce the set of \textit{labels}:
\begin{equation}
\mathfrak{I}=\{\emptyset\} \cup \bigcup_{m=1}^{+\infty} \{0,1\}^m.
\end{equation}
For every $i\in \mathfrak{I}$, there exists a unique $m\in \N^*$ such that $i\in \{0,1\}^m$ and we note $|i|=m$ the generation of individual $i$. For all  $i^1=i^1_1\cdots i^1_{m_1}$ and $i^2=i^2_1\cdots i^2_{m_2}\in \mathfrak{I},$ we define their concatenation  $i^1 i^2$ by the label $i^1_1\cdots i^1_{m_1}i^2_1\cdots i^2_{m_2}.$ We write  $i^1\leq i^2$ when there exists $i^3\in \mathfrak{I}$ such that $i^2=i^1 i^3.$


\par 
Let $V_t\subset \mathfrak{I}$ be the set of cells alive at time $t$ and $N_t=\# V_t $ be its size. For $i\in V_t$ at time $t$, we denote by  $X^i_t\in \R_+$ the quantity of parasites in the cell $i$. The population of cells at time $t$ may be represented by the random point measure on $\mathfrak{I}\times \R_+$, $\bar{Z}_t(du,dx)=\sum_{i\in V_t}\delta_{(i,X_t^i)}(du,dx)$. We define
\begin{equation}Z_t(dx)=\sum_{i\in V_t}\delta_{X_t^i}(dx),\label{defz}
\end{equation}the marginal measure of $\bar{Z}_t(du,dx)$ on the state space $\R_+$.\\
The space of finite measures on $\R_+$, $\mathcal{M}_F(\R_+)$, is embedded with the topology of weak convergence. For a measure $\mu\in \mathcal{M}_F(\R_+)$ and a positive function $f$, we use the notation $\langle \mu, f\rangle=\int_{\R_+} f\, d\mu$.
\par The evolution of (\ref{defz}) can be modelled by a stochastic differential equation (SDE) that is now introduced. This defines a solution in the space $\D(\R_+,\mathcal{M}_F(\R_+))$ of càdlàg measure-valued processes. This space is embedded with the Skorokhod topology (see \textit{e.g.} \cite{joffemetivier}). We follow in this the inspiration of \cite{fourniermeleard, bansayedelmasmarsalletran}.\\

Let $(B^i,i\in \mathfrak{I})$ be a family of independent Brownian motions (BMs) and let $Q(ds,du,di,d\theta)$ be a Poisson point measure (PPM) on $\R_+\times \R_+\times \mathfrak{I}\times [0,1]$ of intensity $q(ds,du,di,d\theta)=ds\, du\,n(di)\,K(d\theta)$ independent from the BMs. We have denoted by $n(di)$ the counting measure on $\mathfrak{I}$ while $ds$ and $du$ are Lebesgue measures on $\R_+$. We denote by $(\mathcal{F}_t : t\geq 0)$ the canonical filtration associated with the BMs and the PPM. Then, for every $(t,x)\rightarrow f(t,x)\in \Co^{1,2}_b(\R_+\times \R_+,\R)$ (bounded of class $\Co^1$ in $t$ and $\Co^2$ in $x$ with bounded derivatives),
 \begin{align}
&\langle Z_t,f\rangle =  f(0,x_0) +\int_0^t\int_{\R_+}\left(\partial_s f(s,x)+gx\partial_x f(s,x)+\sigma^2 x \partial^2_{xx} f(s,x)\right)Z_s(dx)\,ds + M^f_t +\label{martingalegdepop} \\
& \int_0^t \int_{\R_+\times \mathfrak{I}\times [0,1]} \ind_{i\in V_{s_-},\, u\leq r(X^i_{s_-})}\left(f(s,\theta X^i_{s_-})+f(s,(1-\theta)X^i_{s_-})-f(s,X^i_{s_-})\right)Q(ds,du,di,d\theta), \nonumber
\end{align}
where $x_0$ is the quantity of parasite in the ancestor cell $\emptyset$ at $t=0$ and
 \begin{align}
M^f_t=\int_0^t \sum_{i\in V_s}\sqrt{2 \sigma^2 X^i_s}\partial_x f(s,X^i_s)dB^i_s\end{align}is a continuous square integrable martingale with quadratic variation:
\begin{align}
\langle M^{f}\rangle_t= & \int_0^t\int_{\R_+}2\sigma^2 x f^{'2}(s,x)\,Z_s(dx)\,ds.\label{crochetmartingalegdepop}
\end{align}

\begin{prop}
For given initial condition $x_0$, PPM $Q$ and BM $(B^i,\, i\in \mathfrak{I})$, there exists a unique strong solution to (\ref{martingalegdepop})-(\ref{crochetmartingalegdepop}).
\end{prop}

\begin{proof}
Note that as expected, the total quantity of parasites $X_t=\int_{\R_+}x \,Z_t(dx)$ follows the Feller diffusion  (\ref{Fellerdrifted}) with drift $g$ and variance $\sigma^2$. We recall that for this Feller diffusion:
\begin{equation}
\forall t\in \R_+,\, \E(X_t)= x_0 e^{gt}<+\infty.\label{equationmomentfeller}
\end{equation}For existence and uniqueness, we use (\ref{equationmomentfeller}) and refer to similar computation in \cite{champagnatmeleard} and \cite{chithese} Chapter 2 Section 2.2.\\
Notice that for $f\in \Co^{1,2}_b(\R_+^2,\R)$, the process $M^f$ is a real square integrable martingale since for all $t\in \R_+$:
$$2\sigma^2 \|f'\|_\infty^2 \E\Big(\int_0^t \int_{\R_+}x\,Z_s(dx) ds\Big)<+\infty.$$
This ends the proof. \end{proof}

\subsection{From the microscopic discrete model to the continuous model}\label{section_micro}
The continuous model defined above is the limit of a discrete microscopic model where the parasites follow a birth and death process. 
In the microscopic model that we consider, each cell hosts a discrete parasite population. The initial cell contains $[n x_0]$ parasites, where $n\in \N^*$ is a parameter that will tend to infinity and where $[x]$ denotes the integer part of $x$. The parasites are reweighted by $1/n$, so that the biomass in a cell has a constant size order. They reproduce asexually and die with the following respective individual rates:
\begin{align}
b_n=n\sigma^2+b,\quad
d_n=n\sigma^2+d
\end{align}where $b,d>0$ are such that $b-d=g>0$. The cell population is described by the point measure
$$Z^n_t(dx)=\sum_{i\in V_t}\delta_{X_t^{n,i}}(dx)$$
 where $X_t^{n,i}$ is the number of parasites renormalized by $n$ in the cell $i$ at time $t$.

Let $Q^1$ and $Q^2$ be two PPMs on $\R_+\times E_1:=\R_+\times \mathfrak{I}\times \R_+\times \R_+$ and $\R_+\times E_2:=\R_+\times \mathfrak{I}\times \R_+$ with intensity measures $ds \,n(di)\,du$ and $ds \,n(di)\,du\, K(d\theta)$. We associate $Q^1$  to the cell divisions while $Q^2$ corresponds to the births and deaths of the parasites.
\begin{align}
& Z^n_t=  \delta_{(\emptyset,[nx_0]/n)} \nonumber \\
&+ \int_0^t \int_{E_1} Q^1(ds,di,du,d\theta)\ind_{i\in V_{s_-},\ u\leq r(X_{s_-}^{n,i})}\left(\delta_{(i1,[\theta n X_{s_-}^{n,i}]/n)}+\delta_{(i2, X_{s_-}^{n,i}-[\theta n X_{s_-}^{n,i}]/n)}-\delta_{(i,X_{s_-}^{n,i})}\right)\nonumber\\
&+  \int_0^t \int_{E_2} Q^2(ds,di,du)\ind_{i\in V_{s_-}}\left[\left(\delta_{(i,X_{s_-}^{n,i}+1/n)}-\delta_{(i,X_{s_-}^{n,i})}\right)\ind_{u\leq  b_n X^{n,i}_{s_-}}\right.\nonumber\\
  &\qquad \qquad \qquad \qquad \qquad +  \left. \left(\delta_{(i,X_{s_-}^{n,i}-1/n)}-\delta_{(i,X_{s_-}^{n,i})}\right)\ind_{b_n X^{n,i}_{s_-}< u \leq (b_n+d_n)X^{n,i}_{s_-}}\right].\label{eds}
\end{align}
Note that other discrete models would lead to this continuous model. For example, parasites may have several offspring's and their  sharing  in the two daughter cells could follow a random binomial distribution. This means that  we draw $\theta$ in the distribution $K(d\theta)$ and send each parasite in the first daughter cell with the probability $\theta$, and else send it in the second daughter (see also \cite{kimmel, bansaye}).\\
The parasite population is a continuous time birth and death process of constant rates $b_n$ and $d_n$. Hence for every $n\in \N^*$, there is existence and strong uniqueness of the solution of (\ref{eds}) for a given initial condition $Z^n_0$ and PPMs $Q^1$ and $Q^2$.


\begin{prop}\label{propconvergence}
Assume that  there exists an integer $p>0$ and a positive $\bar{r}>0$ such that for all $x\in \R_+$, $0\leq r(x)\leq \bar{r}(1+x^p)$.
Then, the sequence $(Z^n : n\in \N^*)$ defined in (\ref{eds}) converges in distribution in
$\mathbb{D}(\R_+,\mathcal{M}_F(\R_+))$  as  $n\rightarrow +\infty$ to the process
$Z$ defined in (\ref{martingalegdepop})-(\ref{crochetmartingalegdepop}).
\end{prop}
\noindent The proof is deferred in Appendix (Section \ref{appendixmicro}). Additional regularities on the division rate $r$ would be required to control the difference between the microscopic process (\ref{eds}) and its approximation (\ref{martingalegdepop})-(\ref{crochetmartingalegdepop}).

\section{Feller diffusion with multiplicative jumps}\label{critFm}


\par We are interested in the evolution of the quantity of parasites in a  cell line. This means that at each division, we only keep one cell and consider the quantity of parasites inside. This  process $(Y_t : t\geq 0)$   follows
a Feller diffusion with multiplicative jumps whose sizes are independent and  distributed as the r.v. $\Theta$ and which occur at rate $r(x)$ when the process is equal to  $x$.
Without loss of generality, we decide to follow the quantity of parasites in  the 'first branch' of the tree, which is constituted by individuals labeled only with 0s. More precisely, letting
$$\beth=\{\emptyset\} \cup\bigcup_{m\in \N^*}\{0\}^m,$$
this process $(Y_t : t\geq 0)$ is defined by:
$$Y_t=X_t^{i_0(t)}, \quad \t{where} \ i_0(t)\in \beth \cap V_t.$$
Using (\ref{martingalegdepop}), it satisfies
\begin{align}
Y_t 
= &
x_0 +\int_0^t gY_s ds
+\int_0^t \sqrt{2\sigma^2 Y_s}d\beta_s- \int_0^t \int_{\R_+\times [0,1]}
\ind_{u\leq r(Y_{s_-})} \big(1-\theta\big) Y_{s_-}  \rho(ds,du,d\theta)\label{defY}
\end{align}where
\begin{align*}
d\beta_t =\sum_{i\in V_t} \ind_{\beth}(i) dB^i_t\quad \mbox{ and }\rho(ds,du,d\theta)=\ind_{\beth \cap V_{s_-}}(i)Q(ds,du,di,d\theta)
 \end{align*}
are  respectively a BM and  a PPM on  $E:=\R_+\times \R_+\times [0,1]$ with intensity  $ds\,du\,K(d\theta)$. In the sequel, we denote by $(\mathcal{F}^\beta_t : t\geq 0)$ and $(\mathcal{F}^\rho_t : t\geq 0)$ the canonical filtrations associated with the BM $\beta$ and the PPM $\rho$ respectively.\\
Apart from the biological motivations considered here, one can  notice that such Markov processes with multiplicative jumps have various applications (see for instance the TCP in \cite{dgr, grd, jac}).

\subsection{Extinction criterion  when $r$ is constant}\label{sectiontauxconstant}

We first study the asymptotic behavior of the process $Y$ when the jump rate does not depend on the value
of the process.
\begin{prop}\label{extlignee} We assume that $r(.)=r$ is constant. \\
(i) If $g\leq \E(\log(1/\Theta))r$, then  $\P(\exists t >0 : Y_t=0)=1$. \\
Moreover if $g< \E(\log(1/\Theta))r$, 
\be
\label{tempsderetour}
\exists \alpha>0,\, \forall x_0\geq 0,\, \exists c>0,\quad \P_{x_0}(Y_t>0)\leq ce^{-\alpha t} \qquad (t\geq 0).
\ee
(ii) If $g>\E(\log(1/\Theta))r$, then  $\mathbb{P}(\forall t\geq 0 : \  Y_t>0)>0$.\\
Furthermore, for every $0\leq \kappa<g-\E(\log(1/\Theta))r$,
\begin{equation}\{\lim_{t\rightarrow +\infty} e^{-\kappa t}Y_t=\infty\}=\{\forall t: \  Y_t>0 \} \quad \t{a.s.}
\end{equation}
\end{prop}

To guess this extinction criterion one can observe that without division the parasite population follows a Feller diffusion with $\E(X_t)=x_0 \exp(gt)$. With the multiplicative jumps corresponding to the cell divisions, we obtain:
$$Y_t\approx x_0 e^{gt}\prod_{j=1}^{\mathcal{N}_t}\Theta_j \approx x_0\exp\big(gt+\mathcal{N}_t\times \E(\log(\Theta))\big)\qquad (t\rightarrow \infty)$$
where $\mathcal{N}_t$ is a Poisson r.v. of parameter $rt$ and where the $\Theta_j$'s are \iid r.v. with distribution $K(d\theta)$ and independent of $\mathcal{N}_t$. \\
More rigorously, to prove the proposition above, we compute the Laplace transform of the jump-diffusion process (\ref{defY}). Let us introduce the following rescaled process corrected with its drift and jumps:
$$
\bar{Y}_t=  Y_t e^{K_t},
$$with
\be
K_t=-gt-\int_0^t \int_{\R_+\times [0,1]} \ind_{u\leq r} \log(\theta) \rho(ds,du,d\theta)=-gt-\sum_{j=1}^{\mathcal{N}_t}\log(\Theta_j).\label{defKt}
\ee


\begin{lemme}\label{propcorrectionsauts} The process $(\bar{Y}_t : t\geq 0)$ is a continuous local martingale  and for all  $t,\lambda,x_0\geq 0$,
\begin{align}
\label{exposantdeLaplace}
\mathbb{E}_{x_0}\left(\exp(-\lambda \bar{Y}_t)\right)= & \mathbb{E}\left(\exp\left(-\frac{\lambda x_0}{\sigma^2 \lambda \int_0^t e^{K_s}ds+1}\right)\right).
\end{align}
\end{lemme}

\begin{proof}Using that for every $t\in \R_+,\, 0\leq Y_t\leq X_t$ and (\ref{equationmomentfeller}),  all the stochastic integrals that we
are writing  are well defined as local martingales.
Using Itô's formula with jumps (\eg Ikeda Watanabe Th.5.1 on p.67 \cite{ikedawatanabe}) :
\begin{align*}
\bar{Y}_t= & x_0+\int_0^t e^{K_s}\left[gY_s ds+\sqrt{2\sigma^2 Y_s} d\beta_s\right]-  \int_0^t gY_s e^{K_s} ds
\nonumber\\
& \quad +  \int_0^t \int_{\R_+\times [0,1]} \left(Y_s e^{K_s}- Y_{s_-}e^{K_{s_-}}\right)\ind_{u\leq r}\rho(ds,du,d\theta)\nonumber\\
= & x_0+ \int_0^t e^{K_s}\sqrt{2\sigma^2 Y_s}d\beta_s+\int_0^t \int_{\R_+\times [0,1]} \bar{Y}_{s_-}\left(\theta e^{-\log(\theta)}-1\right)\ind_{u\leq r}\rho(ds,du,d\theta)\nonumber\\
= & x_0+ \int_0^t e^{K_s}\sqrt{2\sigma^2 Y_s}d\beta_s.
\end{align*}
Then  $(\bar{Y}_t : t\geq 0)$ is a continuous local martingale which satisfies:
\begin{equation}
\bar{Y}_t=x_0+ \int_0^t e^{K_s/2} \sqrt{2\sigma^2 \bar{Y_s}}d\beta_s.\label{EDSYtilde}
\end{equation}
Let us now work conditionally on $\mathcal{F}_\infty^\rho$. 

\noindent Using Itô's formula for a function $u(t,y)$ which is differentiable by parts with respect to $t$ and infinitely differentiable with respect to $y$, we get
\begin{align*}
 u(t,\bar{Y}_t)= u(0,x_0)+ & \int_0^t \left[\frac{\partial u}{\partial s}(s,\bar{Y}_s)+\frac{\partial^2 u}{\partial y^2}(s,\bar{Y}_s)\sigma^2 \bar{Y}_s e^{K_s}\right]ds  +  \int_0^t \frac{\partial u}{\partial y}(s,\bar{Y}_s)e^{K_s/2}\sqrt{2\sigma^2 \bar{Y}_s}d\beta_s.
\end{align*} Following Ikeda and Watanabe \cite{ikedawatanabe} (8.10) on p.236, we choose  $u(t,y)$ so that the finite variation part equals
zero. More precisely, letting  $t_0\geq 0$ and
\begin{equation}
u(t,y):=\exp\left(-\frac{\lambda y}{\sigma^2 \lambda \int_{t}^{t_0} e^{K_s}ds+1}\right) \quad (0\leq t\leq t_0),
\end{equation} so that
\begin{align*}
u(t,\bar{Y}_t) = u(0,x_0) - \int_0^t \frac{\lambda \sqrt{2\sigma^2 \bar{Y}_s}}{\sigma^2 \lambda \int_{s}^{t_0} e^{K_u}du+1}
\exp\left(\frac{K_s}{2}-\frac{\lambda \bar{Y}_s}{\sigma^2 \lambda \int_{0}^{t-s}e^{K_u}du+1} \right)
d\beta_s.
\end{align*}The process $(u(t,\bar{Y}_t) : 0\leq t\leq t_0)$ is a local martingale bounded by $1$ and thus a real martingale. We  deduce that
$$\mathbb{E}_{x_0}\left(u(t_0,\bar{Y}_{t_0})\, |\, \mathcal{F}^\rho_\infty\right)=\mathbb{E}_{x_0}\left(u(0,\bar{Y}_{0})\, |\, \mathcal{F}^\rho_\infty\right),$$
which gives

\begin{equation}
\mathbb{E}_{x_0}\left(\exp(-\lambda \bar{Y}_{t_0})\, |\, \mathcal{F}^\rho_\infty\right)=\exp\left(-\frac{\lambda x_0}{\sigma^2 \lambda \int_0^{t_0} e^{K_s}ds+1}\right).\label{exposantdeLaplace_pourvince}
\end{equation}
This provides (\ref{exposantdeLaplace}) by taking the expectation.\end{proof}

We can now prove Proposition \ref{extlignee} in the case of  constant division rate.

\begin{proof}[Proof of Proposition \ref{extlignee}]The map $t\mapsto \int_0^t \exp(K_s)ds$ is an a.s. increasing function and thus $$Z_\infty:=\int_0^{+\infty}\exp(K_s)ds\in \R_+\cup \{+\infty\}$$ is well defined. Using dominated convergence, the r.h.s. of (\ref{exposantdeLaplace}) converges to
$$\E_{x_0}(\exp(-\lambda x_0/(\sigma^2 \lambda Z_\infty+1)).$$ Using Lemma \ref{propcorrectionsauts},
the process $(\bar{Y}_t : t\geq 0)$  converges  in distribution as $t\rightarrow + \infty$  to  $\bar{Y}_{\infty}$  whose distribution is specified by
\begin{equation}
\E_{x_0}(e^{-\lambda \bar{Y}_{\infty}})=\E_{x_0} \left(\exp\left(-\frac{\lambda x_0}{\sigma^2 \lambda  Z_{\infty}+1} \right)\right).
\label{ybarinfty_Laplace}
\end{equation}
Recalling from (\ref{EDSYtilde}) that $(\bar{Y}_t : t\geq 0)$ is a positive local martingale, we obtain by Jensen's inequality that $(\exp(-\bar{Y}_t) : t\geq 0)$ is a positive sub-martingale bounded by 1. From this, we deduce that the convergence towards $\bar{Y}_\infty$, which is possibly infinite, also holds a.s.\\
Letting  $\lambda\rightarrow +\infty$, we get by bounded convergence:
\begin{align}
\label{pext}
\P_{x_0}(\bar{Y}_{\infty}=0)= & \lim_{\lambda\rightarrow +\infty}\E_{x_0}\left(\exp\left(-\frac{\lambda x_0}{\sigma^2 \lambda Z_{\infty}+1}\right)\right)
=  \mathbb{E}_{x_0}\left(\exp\left(-\frac{x_0}{\sigma^2 Z_{\infty}}\right)\right).
\end{align}
From \cite{bertoinlevyprocesses} Corollary 2 p.190, we know that the Lévy process $(K_t : t\geq 0)$ defined in (\ref{defKt})
tends to $+\infty$ (resp. $-\infty$, resp. it oscillates) when $\E(K_1)=\E(\log(1/\Theta))r-g$ is positive (resp. negative, resp. zero). Thus, we split the proof into three cases.\\

If \underline{$g>\E(\log(1/\Theta))r$}, we  choose $\kappa>0$ such that  $g -\kappa>\E(\log(1/\Theta))r$. Then $(K_t+\kappa t : t\geq 0)$ is a Lévy process
such that $\E(K_1+\kappa)<0$, so
\begin{equation*}
\lim_{t\rightarrow +\infty}K_t+\kappa t= -\infty\quad \mbox{ a.s.}
\end{equation*}
We define $B:=\sup_{t\geq 0}\{K_t+\kappa t\}<\infty$ a.s. so that $K_t \leq -\kappa t +B$ a.s.
This ensures that  $Z_{\infty}<\infty$ a.s. and using (\ref{pext}), we get:
\begin{equation}
\P(\bar{Y}_{\infty}=0)<1.\label{etape8}
\end{equation}
As $Y_t=\bar{Y}_t \exp(-K_t)$ and $\lim_{t\rightarrow +\infty}\exp(-K_t)=+\infty,$ we have:
\begin{equation*}
\mathbb{P}\left(\forall t>0: \ Y_t>0\right)\geq \mathbb{P}\left(\lim_{t\rightarrow +\infty}Y_t =+\infty\right)\geq \mathbb{P}\left(\lim_{t\rightarrow +\infty}\bar{Y}_t > 0\right)>0.
\end{equation*}
Furthermore, for every $0\leq \kappa<g-\E(\log(1/\Theta))r$, $-K_t-\kappa t\rightarrow + \infty$ as $t\rightarrow \infty$. So
$A=\{ \lim_{t\rightarrow +\infty}K_t+\kappa t=+\infty\}\in \mathcal{F}^\rho_\infty$ happens a.s. and
\begin{equation*}
\P\left(\lim_{t\rightarrow +\infty} e^{-\kappa t}Y_t=+\infty \ \vert \ \mathcal{F}^\rho_\infty\right)\geq \mathbb{P}\left(\lim_{t\rightarrow +\infty}\bar{Y}_t > 0  \ \vert \ \mathcal{F}^\rho_\infty  \right)>0\ a.s.
\end{equation*}
by using (\ref{exposantdeLaplace_pourvince}) as in (\ref{etape8}). We can  now  prove that $\{\lim_{t\rightarrow +\infty} e^{-\kappa t}Y_t=+\infty\}=\{\lim_{t\rightarrow +\infty}Y_t =+\infty \}$ a.s.
Let $N>0$. Conditionally on the event $\{Y_t\rightarrow \infty\}$, the stopping time $T_N=\inf\{t \geq 0 : Y_t\geq N\}$ is finite for every $N$. Since
$\P_1\left(\lim_{t\rightarrow +\infty} e^{-\kappa t}Y_t=+\infty \ \vert \ \mathcal{F}^\rho_\infty\right) >0$ a.s., and since the process $(Y_t : t\geq 0)$ satisfies the branching property conditionally to $\mathcal{F}^\rho_\infty$, this ensures that
$$\P_N\left(\lim_{t\rightarrow +\infty} e^{-\kappa t}Y_t<+\infty \ \vert \ \mathcal{F}^\rho_\infty\right)\leq \P_1\left(\lim_{t\rightarrow +\infty} e^{-\kappa t}Y_t<+\infty \ \vert \ \mathcal{F}^\rho_\infty\right)^N \rightarrow 0$$
as $N\rightarrow + \infty$. Thus  conditionally on  $\mathcal{F}^\rho_\infty$ and $\{Y_t\rightarrow +\infty\}$, $e^{-\kappa t}Y_t\rightarrow +\infty$ a.s. Finally the fact that $\{\lim_{t\rightarrow +\infty}Y_t =+\infty \}=\{\forall t: \  Y_t>0 \}$ a.s. is a classical consequence of the Markov property using that $0$ is an absorbing state.
\\

If \underline{$g=\E(\log(1/\Theta))r$}, the Lévy process $(K_t : t\geq 0)$ oscillates a.s.:
\begin{equation}
\limsup_{t\rightarrow +\infty} K_t= +\infty, \qquad \liminf_{t\rightarrow +\infty} K_t=-\infty.\label{defoscillation}
\end{equation}
Then, for every  $k\geq 0$, the stopping time $T_k:=\inf \{t\geq 0 : K_t\geq k\}$ is finite a.s. This ensures that almost surely
\begin{align}
Z_{\infty} \geq & \int_{T_k}^{T_k+1} e^{K_t} \d t \geq e^k \int_{T_k}^{T_k+1} e^{K_t-K_{T_k}} \d t \geq  e^k \exp\big(\inf_{t\in [T_k,T_k+1]} \{K_t-K_{T_k}\}\big).\label{etape3}
\end{align}
As  $(\inf_{t\in [T_k,T_k+1]}\{K_t-K_{T_k}\} : k \geq 1)$ are identically distributed finite r.v., we have
$$\limsup_{k\in \N^*}e^k \exp\big(\inf_{t\in [T_k,T_k+1]} \{K_t-K_{T_k}\}\big)=+\infty\quad \mbox{ a.s.}$$
Since the \lhs of (\ref{etape3}) does not depend on $k$, letting $k\rightarrow \infty$ ensures that $Z_{\infty}=+\infty \quad \t{a.s.}$ and (\ref{pext}) gives:
\begin{equation}\bar{Y}_{\infty}=0 \quad \t{a.s.} \label{etapeoscillation}\end{equation}

Our purpose is now to prove that $(Y_t : t\geq 0)$ reaches $0$ in finite time \as Let us define the following stopping times, which are finite from (\ref{defoscillation}) and (\ref{etapeoscillation}):
\begin{equation}
\tau_1:=\inf\{t\geq 0 : Y_t\leq 1\}, \qquad \tau_{i+1}:=\inf\{t >\tau_i+1 : Y_t\leq 1\}  \ \ (i\geq 1).\label{tempsdarrettau}
\end{equation}
Introducing $a:= \inf_{x\in [0,1]} \{\P_x(Y_1=0)\}>0$ we have:
\bea
\P( Y_{\tau_j}>0)&= &\P(Y_{\tau_1}>0)\prod_{i=2}^j \P(Y_{\tau_{i}}>0 \ \vert \ Y_{\tau_{i-1}}>0) \nonumber\\
&\leq & \P(Y_{\tau_1}>0)\prod_{i=2}^j \P(Y_{\tau_{i-1}+1}>0 \ \vert \ Y_{\tau_{i-1}}>0)
\leq  \P(Y_{\tau_1}>0)(1-a)^{j-1},\label{utptemarkov}
\eea
which tends to $0$ as $j\rightarrow \infty$. Then, $\P(\exists n \in \N : Y_{\tau_n}=0)=1,$ which is the desired result.\\

Finally if \underline{$g<\E(\log(1/\Theta)) r$}, we choose $\kappa>0$ so that $g +\kappa<\E(\log(1/\Theta)) r$. Then $K_t-\kappa t\rightarrow +\infty$ a.s. Proceeding as in the case $g>\E(\log(1/\Theta))r$ with $B:=\inf_{t\in \R_+}\{K_t-\kappa t\}>-\infty$ a.s., we obtain $K_t\geq \kappa t +B$ a.s. This implies that $Z_{\infty}=+\infty$ \as Using (\ref{pext}), we get
$\bar{Y}_{\infty}=0 \quad \t{a.s.}$ Moreover  $\exp(-K_t)\rightarrow 0$ as $t\rightarrow \infty$ a.s., so
\begin{equation}
\lim_{t\rightarrow +\infty}Y_t=  \lim_{t\rightarrow +\infty}\bar{Y}_t e^{-K_t}= 0 \quad \t{a.s.}
\end{equation}
Using again the stopping times  (\ref{tempsdarrettau}),  we get  that $Y_t$ reaches $0$ in finite time a.s.\\

Let us now prove (\ref{tempsderetour}). Formula (\ref{exposantdeLaplace}) yields :
$$\P_{x_0}(\bar{Y}_t=0)=\lim_{\lambda\rightarrow +\infty} \mathbb{E}\left(\exp\left(-\frac{\lambda x_0}{\sigma^2 \lambda \int_0^t e^{K_s}ds+1}\right)\right)= \mathbb{E}\left(\exp\left(-\frac{ x_0}{\sigma^2  \int_0^t e^{K_s}ds}\right)\right).$$
As the process $(K_t :t\geq 0)$ has no negative jumps and drift $-g$, we have
$$\inf_{u\in [t-1,t]} K_u\geq K_{t-1}-g\qquad \mbox{\as}$$
for every $t\geq 1$  and
$$\int_0^t e^{K_s}ds\geq \int_{t-1}^t e^{K_s}ds \geq e^{K_{t-1}-g}\qquad \mbox{\as}$$
Moreover for all $x\geq 0$ and $\alpha \in [0,1]$, $1-e^{-x}\leq \min(1,x)\leq \min(1,x^{\alpha})\leq x^{\alpha}$.
This  gives for every $t\geq 0$,
\Bea
\P_{x_0}(\bar{Y}_t>0)&=& \E\left(1-\exp\left(-x_0\sigma^{-2} e^{-K_{t-1}+g} \right)\right)\\
&\leq & \big( x_0\sigma^{-2}  e^{g}\big)^{\alpha} \mathbb{E}\left(e^{-\alpha K_{t-1}} \right) \\
&=& \big(x_0\sigma^{-2}  e^{g}\big)^{\alpha}e^{-(t-1)\phi(\alpha)},
\Eea
using the Lévy-Khintchine formula where $\phi$ is the Laplace exponent of $(K_t)_{t\geq 0}$ (see \cite{bertoinlevyprocesses}):
$$\phi(\alpha):=  -g\alpha+ r \E(1-e^{-\alpha \log(1/\Theta)}) \qquad (\alpha\geq 0).$$
Adding that $\phi(0)=0$ and $\phi'(0)=r\E(\log(1/\Theta))-g>0$, there exists $\alpha\in(0,1]$ such that $\phi(\alpha)>0$ and:
$$\P_{x_0}(Y_t>0)=\P_{x_0}(\bar{Y}_t>0) \leq \big[\big(x_0\sigma^{-2}  e^{g}\big)^{\alpha}e^{\phi(\alpha)}\big] e^{- t \phi(\alpha)},$$
which completes the proof.\end{proof}

\subsection{Extinction criteria  with monotone division rate}\label{sectiontauxnonconstant}

$\qquad$ We give here the extinction criteria of the process $(Y_t : t\geq 0)$ describing the quantity of parasites in a cell line when the jump rate $r$ is monotone. For the proof,
we  use  coupling arguments to compare this process with the case of  constant division rate. \\

We begin with the case where $r$ is an increasing function which means that the more parasites the cells contain, the faster
they divide. This case is relevant when the cell division rate is increased to get ride of the parasites
or   when there is a symbiosis between parasites and cells. The asymptotic behavior of $Y$ depends on the maximum division rate
\begin{equation}
r^*:=\sup_{x\in\R^+}r(x).
\label{defrstar}
\end{equation}

\begin{prop}\label{linerincre} We assume that $r$ is an increasing function. \\
(i) If there exists  $x_1\geq 0$ such that  $g\leq \E(\log(1/\Theta))r(x_1)$, then
$$\P\Big(\exists \ t>0, \ Y_t=0\Big)=1.$$
(ii) If $g>\E(\log(1/\Theta))r^*$,   then $\P(\forall t\geq 0: \  Y_t>0)>0$.\\ Furthermore, for every
$0\leq \kappa<g-\E(\log(1/\Theta))r^*,$ we have a.s.
$$\{\lim_{t\rightarrow +\infty}e^{-\kappa t}Y_t=\infty\}=\{\forall t\geq 0: \  Y_t>0\}.$$
\end{prop}
Let us note that  the case $g>\E(\log(1/\Theta))r(x)$ for every $x\geq 0$ and $g=\E(\log(1/\Theta))r^*$ remains open. The expected result is a.s. extinction but this may depend on the speed of convergence of $r(x)$ to $r^*$ as $x\rightarrow \infty$.

\begin{proof} Heuristically, if $g\leq \E(\log(1/\Theta))r(x_1)$, as soon as $Y\geq x_1$, the division rate is larger than $r(x_1)$ and
Proposition \ref{extlignee} ensures that the process is pushed back to $x_1$. Eventually, it reaches zero. \\
We give the proof of $(i)$ using a coupling argument. Let us define $\w{Y}$ as:
\begin{align*}
\w{Y}_t = &
Y_1 +\int_0^{t} g \w{Y}_{s} ds
+\int_1^{t+1} \sqrt{2\sigma^2 \w{Y}_{s-1}}dB_s^0- \int_1^{t+1} \int_{\R_+\times [0,1]}
\ind_{u\leq r(x_1)} \theta \w{Y}_{s_- -1}  \rho(ds,du,d\theta)
\end{align*}
with initial condition $Y_1$, with the same BM $\beta$ and PPM $\rho$ as $Y$
shifted from $1$, and with the constant division rate $\w{r}(x)=r(x_1)$.  \\
 Thus, $\w{Y}$ is a Feller diffusion with multiplicative jumps given by $\Theta$ and constant division rate $r(x_1)$. Proposition
\ref{extlignee} ensures that it becomes extincted in  finite time. \\
Moreover, denoting by
$$\tau_1:=\inf\{t\geq 1 : Y_t\leq x_1\} \in [0,\infty],$$
 the definition of $\w{Y}$ ensures that
 $$Y_{1+t}\leq \w{Y}_t \quad a.s., \qquad (0\leq t\leq \tau_1),$$
 since $\w{Y}$ undergoes less jumps than $Y_{1+.}$. As $\w{Y}$ becomes extincted in finite time, $\tau_1<\infty$ a.s. \\
Similarly, the following stopping times:
\begin{equation}
\tau_0:=0, \quad \tau_{i+1}:=\inf\{t\geq \tau_i +1 : Y_t\leq x_1\}
\end{equation}
are finite a.s. for $i\in\N$.
Proceeding as in (\ref{utptemarkov}), we obtain that $\P(\exists i\in \N : Y_{\tau_i}=0)=1$. This ends the proof of (i).\\


We now prove $(ii)$ and we assume that $g>\E(\log(1/\Theta))r^*.$ We define $(\w{Y}_t : t\geq 0)$ with the same BM and PPM as $(Y_t : t\geq 0)$ except that the indicator in (\ref{defY}) is replace by $\ind_{u\leq r^*}$. The definition of $r^*$ and this pathwise construction ensure that for every
$t\geq 0$, $Y_t\geq \w{Y}_t$ a.s. Moreover $\w{Y}$ is a Feller diffusion with multiplicative jumps with constant rate $r^*$ and Proposition \ref{extlignee} (ii) states that $\w{Y}$ grows geometrically with positive probability.  Thus the same holds for
 $Y$. Combining the Markov property with Proposition \ref{extlignee} (ii) ensures that
$\{\lim_{t\rightarrow +\infty} \exp(-\kappa t) Y_t \}
= \{ \forall t\geq 0: \  Y_t>0\}$ a.s. since we have the analogous result  for the coupling process $\w{Y}$.
\end{proof}

We now consider the case where the more parasites there are, the less the cells divide. This is natural
if the parasites make the cells ill or use their nutriments. The asymptotic behavior now depends on
\begin{equation}r_*=\inf_{x\geq 0}r(x).\label{defr_star(inf)}\end{equation}
\begin{prop}\label{extrdecr} We assume that $r$ is a decreasing function. \\
(i) If $g\leq \E(\log(1/\Theta))r_*$, then
$\P\Big(\exists  t>0, \ Y_t=0\Big)=1.$\\
(ii) Else, $\P(\forall t>0: \ Y_t>0)>0$ and for every $0<\kappa<\E(\log(1/\Theta))r_*-g$, we have a.s.
$$ \{\lim_{t\rightarrow \infty} e^{-\kappa t} Y_t= \infty\}=\{ \forall t>0: \ Y_t>0\}).$$
\end{prop}

\begin{proof}
Let us define the process $(\w{Y}_t : t\geq 0)$ with the same BM and PPM as $(Y_t : t\geq 0)$ but we replace the indicator in (\ref{defY}) by $\ind_{u\leq r_*}$. Then, for every  $t\geq 0$, $\w{Y}_t\geq Y_t$ \as and $\w{Y}$ is a Feller diffusion with multiplicative jumps
$\Theta$ and constant jump rate $r_{*}$.  If  $g\leq \E(\log(1/\Theta))r_*$, Proposition \ref{extlignee} ensures that $\w{Y}$ becomes extincted a.s., which entails (i).\\

\par For (ii), and let us consider $x_1$ such that  $g> \E(\log(1/\Theta))r(x_1)$. We define the process $(\w{Y}_t : t\geq 0)$ with the same BM and PPM as $(Y_t : t\geq 0)$ but we replace the indicator in (\ref{defY}) by $\ind_{u\leq r(x_1)}$, so that it divides with the constant rate $r(x_1)$.
Using the Markov property and Proposition \ref{extlignee}, we get
\begin{align*}
\P(\lim_{t\rightarrow \infty} Y_t=\infty)
\geq & \P(Y_1\geq x_1+1)\P_{x_1+1}(\lim_{t\rightarrow \infty} Y_t=\infty) \\
\geq & \P(Y_1\geq x_1+1)\P_{x_1+1}(\lim_{t\rightarrow \infty} \w{Y}_t=\infty; \ \forall t\geq 0,  \ \w{Y}_t\geq x_1) >0,
\end{align*}
which gives the first part of $(ii)$. The second part comes from the Markov property and this coupling argument as for the two previous propositions.
\end{proof}

\section{Recovery criterion for constant division rate}\label{section3}
In this section, we want to determine how the infection evolves in the cell population. More precisely, we are interested in the asymptotic proportions of cells which contain a given quantity of parasites. \\
The questions we focus on do not need spatial structure on the cell population so without loss of generality we assume by now that $\Theta$ is symmetric in distribution with respect $1/2$:
$$\Theta\stackrel{d}{=}1-\Theta.$$
In the case of a constant division rate $r$, $(N_t : t\geq 0)$ is a Yule process and  $\E(N_t)=e^{rt}$. For the recovery criterion of the organism, we are interested in the asymptotic behavior of
$$ Z_t(dx)/N_t.$$
But, as usual for branching Markov process \cite{bansayedelmasmarsalletran, evanssteinsaltz}, it is more convenient to consider the following renormalization :
\begin{equation*}
\gamma_t(dx) := Z_t(dx)/\E(N_t)=e^{-rt}Z_t(dx).\label{defgammat}
\end{equation*}
Actually, we just need here the expectation of this quantity, whose evolution is given by the following result
\begin{lemme}\label{propprocessusauxiliaire2}The family of probability measures $(\E(\gamma_t) : t\geq 0)$ is the unique solution of the following equation in $(\nu_t : t\geq 0)$ for $(f : (t,x)\mapsto f_t(x))\in \Co^{1,2}_b(\R_+\times\R_+,\R)$ and $t\in \R_+$:
\begin{multline}
\langle \nu_t,f_t\rangle
= f_0(x_0) + 2r \int_0^t \int_{\R_+}\int_0^1 \left[f_s(\theta x)-f_s(x)\right] K(d\theta) \,\nu_s(dx)\, ds\\
  + \int_0^t \int_{\R} \left(\partial_sf_s(x)+gx \partial_x f_s(x)+\sigma^2 x \partial^2_{xx}f(x)\right) \nu_s(dx)\, ds. \qquad  \label{nugaltonwatson}
\end{multline}
\end{lemme}
This result ca  be derived directly
from the measure-valued equation (\ref{martingalegdepop}) and we give the proof below.  We can then interpret $\E(\gamma_t)$ as the marginal of an auxiliary process  $(\xi_t : t\geq 0)$:
\be
\forall f\in \Co^{2}_b(\R_+,\R),\qquad \forall t\in \R_+,\, \langle \E(\gamma_t),f\rangle=e^{-rt}\E\big(\sum_{i\in V_t} f(X_t^i) \big)= \mathbb{E}(f(\xi_t)),
\label{relationdelmasmarsalle}
\ee
where $(\xi_t : t\geq 0)$ is a Feller diffusion which jumps with rate $2r$ from $x$ to $\Theta x$. More precisely, it is defined for $t\geq 0$ by
\begin{multline}
\xi_t=  x_0+  \int_0^t g\xi_s ds+\int_0^t \sqrt{2 \sigma^2 \xi_s}dW_s
+\int_0^t\int_{\R \times [0,1]} \ind_{u \leq 2r} [f(\theta \xi_{s_-})-f(\xi_{s_-})]
  N(ds,du,d\theta)\label{definitionxi}
\end{multline}
where $N(ds,du,d\theta)$ is a PPM of intensity $ds\,du\,d\theta$ and $W$, a standard BM independent from $N$. \\

This result  is generalized  for  Markov processes indexed by Galton-Watson trees in \cite{bansayedelmasmarsalletran}, with a different approach which leads to  pathwise representation. 
The auxiliary process jumps with rate $2r$ whereas the cell divides with rate $r$. This bias phenomenon is classical
and have  been obtained in \cite{chauvinrouaultwakolbinger, hardyharris3} with different approaches. It corresponds to the fact that the faster the cells divide, the more descendants they have  at time $t$. That is why the ancestral lineages
from typical individual at time $t$ have an accelerated rate of division $2r$. \\

Using dominated convergence in the l.h.s. and r.h.s. of (\ref{relationdelmasmarsalle}), we show that (\ref{relationdelmasmarsalle}) also holds for $f(x)=\ind_{x>0}$. By Proposition \ref{extlignee}, we can now determine the evolution of parasites in the cell population and prove the recovery criterion, when the division rate is constant.

\begin{theorem}\label{threcoveryorganismrconstant}
(i) If $g\leq  2r\E(\log(1/\Theta))$, then the organism recovers a.s. in the sense that:
$$ \lim_{t\rightarrow +\infty}\frac{\# \{ i \in V_t : X^i_t=0\}}{N_t} = 1,\quad \mbox{ a.s}.$$
(ii) If $g>  2r\E(\log(1/\Theta))$ then the parasites proliferate inside the cells as soon as  the parasites do not become
extinct in the sense that
\begin{equation}
\big\{\limsup_{t\rightarrow \infty} \frac{\# \{ i \in V_t : X^i_t\geq e^{\kappa t} \}}{N_t} > 0 \big\}=\{\forall  t>0: \ X_t>0\} \quad \mbox{a.s.}\label{recoveryrctcroissexp}
\end{equation}
for every  $0\leq \kappa<g- 2r \E(\log(1/\Theta))$. The probability of these events equals $1-\exp(-gx_0/\sigma^2)$.
\end{theorem}
The factor $2$ in the criterion comes from the auxiliary process and 'increases recovery' in the sense that
the quantity of parasites in a cell line $Y$ may go to infinity with positive probability whereas the organism recovers a.s.
Note that there is a zero-one law: either the cells recover or parasites proliferate inside a positive proportion of cells. This dichotomy may fail when $r$ will be an increasing function of the quantity of parasites (Section \ref{sectiontauxnonconstant}).

\begin{proof}[Proof of Lemma  \ref{propprocessusauxiliaire2}]Let $t\in \R_+$ and $(f\:(s,x)\mapsto f_s(x))\in \Co_b^{1,2}(\R_+\times \R_+,\R)$.
Using (\ref{martingalegdepop}) with $(s,x)\mapsto f_s(x)e^{-rs}$ entails:
\begin{multline*}
\langle \gamma_t, f_t\rangle  = f_0(x_0)
 +  \int_0^t \int_{\R} \left(gx \partial_x f_s(x)+\sigma^2 x \partial^2_{xx}f_s(x)-rf_s(x)+\partial_s f_s(x)\right)e^{-rs} Z_s(dx)\, ds + \widetilde{M}_t^f\\
+ \int_0^t \int_{\mathfrak{I}\times \R\times [0,1]}\ind_{i\in V_{s_-}} \ind_{u\leq r}  \left[f_s(\theta X^i_{s-})+f_s((1-\theta)X^i_{s-})-f_s(X^i_{s-})\right]e^{-rs} Q(ds,du,di,d\theta)
\end{multline*}where $\widetilde{M}_t^f$ is a continuous square integrable martingale started at 0. Taking the expectation:
\begin{align}
\langle \E(\gamma_t), f_t\rangle 
= & f_0(x_0) + \int_0^t \int_{\R} \left(gx \,\partial_x f_s(x)+\sigma^2 x \,\partial^2_{xx}f_s(x)+\partial_s f_s(x)\right) \E(\gamma_s)(dx) \, ds \nonumber\\
 & +\int_0^t\int_{\R_+}\int_0^1 2r\left[f_s(\theta x)-f_s(x)\right] K(d\theta) \,\E(\gamma_s)(dx)\, ds,
\label{equationaux}
\end{align}by using the symmetry of $K(d\theta)$.

\par Let us prove that there is a unique solution to (\ref{nugaltonwatson}). Let $(\nu^1_t : t\geq 0)$ and $(\nu^2_t : t\geq 0)$ be two solutions. Recall (\eg \cite{rachev}) that the total variation distance between $\nu^1_t$ and $\nu^2_t$ is
\begin{equation}
\| \nu^1_t-\nu^2_t\|_{TV}=\sup_{\substack{\phi\in \Co_b(\R_+,\R) \\ \|\varphi\|_\infty\leq 1}}|\langle \nu^1_t,\phi\rangle-\langle \nu^2_t,\phi\rangle|.\label{totalvariation}
\end{equation}
Let $t\in \R_+$ and $\varphi\in \Co^2_b(\R_+,\R)$ with $\|\varphi\|_\infty\leq 1$. We denote by $(P_s : s\geq 0)$ the semi-group associated with the Feller diffusion (\ref{Fellerdrifted}) started at $x\in \R_+$: $P_s\varphi(x)=\E_x(\varphi(X_s))$. Notice that $\|P_{t-s}\varphi\|_\infty\leq \|\varphi\|_\infty\leq 1$. Then using (\ref{equationaux}) with $f_s(x)=P_{t-s}\varphi(x)$, the second term equals $0$:
\begin{align}
\left| \langle \nu^1_t-\nu^2_t, \varphi\rangle\right| = & \left|2r \int_0^t\int_{\R_+} \int_0^1  \Big(P_{t-s}\varphi(\theta x)-P_{t-s}\varphi(x)\Big)K(d\theta) (\nu_s^1-\nu^2_s)(dx)\, ds\right|\nonumber\\
\leq & 4r \int_0^t \|\nu^1_s-\nu^2_s\|_{TV} ds.
 \end{align}Since $\Co^2_b(\R_+,\R)$ is dense in $\Co_b(\R_+,\R)$ for the bounded pointwise topology, taking the supremum in the l.h.s. implies that: $\|\nu^1_t-\nu^2_t\|_{TV}\leq 4r \int_0^t \|\nu^1_s-\nu^2_s\|_{TV} ds$. Gronwall's lemma implies  that $\|\nu^1_t-\nu^2_t\|_{TV}=0$.
\end{proof}

For the proof of the theorem, the following Lemma will be used to obtain the almost sure convergence from the convergence in probability:
\begin{lemme} \label{lemtechnq}Let $V$ be a denumerable subset and
 $(N_t(i) : t\geq 0)$  be \iid processes distributed as the Yule process $(N_t : t\geq 0)$ for $i\in V$. Then there
exists a nonnegative nonincreasing function  $G$  on $\R_+$
such that $G(y)\rightarrow 0$ as $y\rightarrow \infty$ and  for all  $I,J$  finite subsets of $V$ and $x\geq 0$:
\begin{equation}
\P\Big(\sup_{t\geq 0}\frac{\sum_{i\in J} N_t(i)}{\sum_{i\in I} N_t(i)} \geq x\Big)\leq G\big(\frac{\#I}{\#J}  x\big).
\end{equation}
\end{lemme}
\begin{proof}
We introduce for every $i\in V$,
$$M(i):=\sup_{t\geq 0} N_t(i)e^{-rt}\quad \mbox{ and } \quad m(i)=\inf_{t\geq 0} N_t(i)e^{-rt}.$$
As $(N_t : t\geq 0)$ is a Yule process,  $(M(i) : i \in V)$ and $(m(i): i \in V)$ are both finite positive \iid
r.v.'s with finite expectation. Moreover
\Bea
\frac{\sum_{i\in J} N_t(i)}{\sum_{i\in I} N_t(i)}&\leq& \frac{\sum_{i\in J} M(i)}{\sum_{i\in I} m(i)}
\leq  \frac{\#J}{\#I}\frac{\sum_{i\in J} M(i)}{\# J} \frac{\# I}{\sum_{i\in I} m(i)}.
\Eea
and the result follows by defining for $y\geq 0$
$$G(y)=\sup \Big\{  \P\Big(\frac{\sum_{i\in J} M(i)}{\# J} \frac{\# I}{\sum_{i\in I} m(i)}\geq y\Big)  : I,J\subset V; \#I,\#J < \infty\Big\}.$$Indeed,  by the law of large numbers, the sequence
$$\frac{\sum_{i\in J} M(i)}{\# J} \frac{\# I}{\sum_{i\in I} m(i)}$$
is uniformly tight. So $G(y)\rightarrow  0$ as $y\rightarrow \infty$.
\end{proof}

\begin{proof}[Proof of Theorem \ref{threcoveryorganismrconstant}] Let us start with the proof of (i). We can apply (\ref{relationdelmasmarsalle}) with  $f(x)=\ind_{x>0}$ since this function is the increasing limit of function which belong to $f\in \Co^{2}_b(\R_+,\R)$. Then
\begin{equation} \E\left(\frac{\# \{ i \in V_t : X^i_t>0\}}{\E(N_t)}\right)=\E\big(\langle \gamma_t,\ind_{x> 0}\rangle \big) =\P(\xi_t>0),\label{etape4}
\end{equation}
where $(\xi_t: t \geq 0)$ is defined in (\ref{definitionxi}). By Proposition \ref{extlignee}, under the assumption of (i), $(\xi_t : t\geq 0)$ becomes extincted in finite time a.s. so that, $\langle \gamma_t,\ind_{x>0}\rangle$ converges in $L^1$ and hence in probability to $0$. As $\E(N_t)/N_t$ tends in probability to $1/W$ where $W$ is an exponential r.v. of parameter $1$ (\eg Athreya and Ney \cite{athreyaney} Chap.III Sect.4), then
\begin{equation}
\lim_{t\rightarrow +\infty}\frac{\# \{ i \in V_t : X^i_t>0\} }{N_t}=\frac{0}{W}=0\quad \mbox{in probability}.\label{etape5}\end{equation}
To get that the convergence holds a.s. and complete the proof of (i), let us denote by $V_t^*=\{i \in V_t : X^i_t>0\}$ the set of infected cells and by $N_t^*=\# V_t^*$ its cardinal. Denoting by $V_{t,s}(i)$ the set of cell alive at time $t+s$ whose ancestor at time $t$ is the cell $i\in V_t$, we have
$$\frac{N^*_{t+s}}{N_{t+s}} \leq \frac{\sum_{i\in V_t^*} \# V_{t,s}(i)}{\sum_{i\in V_t} \# V_{t,s}(i)},$$
where $(V_{t,s}(i) :s\geq 0)$ are \iid for $i\in V_t$.  As $N_t\rightarrow \infty$ and $N_t^*/N_t\rightarrow 0$ as $t\rightarrow \infty$, we get by Lemma \ref{lemtechnq} that
$$\lim_{t\rightarrow +\infty}\sup_{s\geq 0}\frac{\sum_{i\in V_t^*} \# V_{t,s}(i)}{\sum_{i\in V_t} \# V_{t,s}(i)}=0\qquad \mbox{ in probability.}$$Moreover for all $\epsilon,\eta>0$, there exists $t>0$ such that $\P(N_t^*/N_t\geq \eta)\leq \epsilon.$
Then, choosing $\eta$ small enough, we get that
$$\P(\sup_{s\geq 0}N^*_{t+s}/N_{t+s} \geq  2\epsilon)\leq \P\bigg( \sup_{s\geq 0} \frac{\sum_{i\in V_t^*} \# V_{t,s}(i)}{\sum_{i\in V_t} \# V_{t,s}(i)}\geq 2\epsilon\bigg) \leq 2\epsilon.$$This gives the a.s. convergence and ends up the proof of (i).\\

\par Let us now prove (ii). If there exists $\kappa \in [0, g- 2r \E(\log(1/\Theta)))$
 such that:
\begin{equation}
\P\big(\limsup_{t\rightarrow \infty} \frac{\# \{ i \in V_t : X^i_t\geq \exp(\kappa t)\}}{N_t} > 0 \big)=0.\label{hypabsurde}
\end{equation}
then $\lim_{t\rightarrow +\infty}\# \{ i \in V_t : X^i_t\geq \exp(\kappa t)\}/N_t=0$ in probability. Since $N_t/\E(N_t)$ converges in probability to an exponential \rv $W$ of parameter 1, then
\be
\label{expression}
\lim_{t\rightarrow +\infty}\frac{\# \{ i \in V_t : X^i_t\geq \exp(\kappa t)\}}{\E(N_t)}=0 \quad \mbox{in probability}.
\ee
Moreover
$$\frac{\# \{ i \in V_t : X^i_t\geq \exp(\kappa t)\}}{\E(N_t)}\leq \frac{ N_t }{\E(N_t)},$$
which is bounded in $L^2$. Then
$\langle \gamma_t, \ind_{x\geq \exp(\kappa t)}\rangle=\# \{ i \in V_t : X^i_t\geq \exp(\kappa t)\}/\E(N_t)$ is uniformly integrable and  the convergence in probability of (\ref{expression}) implies the $L^1$ convergence. Thus,
$$\lim_{t\rightarrow +\infty}\P(\xi_t\geq \exp(\kappa t))=\lim_{t\rightarrow +\infty}\E\left(\langle \gamma_t, \ind_{x\geq \exp(\kappa t)}\rangle\right)= 0,$$
which is in contradiction with Proposition \ref{extlignee} (ii). Then (\ref{hypabsurde})  does not hold and  for every $\kappa \in [0, g- 2r \E(\log(1/\Theta)))$,
$$\P\big(\limsup_{t\rightarrow \infty} \frac{\# \{ i \in V_t : X^i_t\geq \exp(\kappa t)\}}{N_t} > 0 \big)>0.$$
Note that $\P(\forall t>0 : X_t>0)>0$ and by a zero one law argument, we prove now that
\be
\label{evnonext}\big\{\limsup_{t\rightarrow \infty} \frac{\# \{ i \in V_t : X^i_t\geq \exp(\kappa t)\}}{N_t} >0\big\}=\{\forall t>0, X_t>0\} \quad \t{a.s.}
\ee
In that view, let us define
$$ V_t^1=\{ i \in V_t : X_i^t\geq 1\}$$
the set of cells at time $t$ whose quantity of parasites is more   than $1$. Let us note that $g>2\E(\log(1/F))r\geq r$ so the exponential growth of parasites is larger than the exponential growth of the number of cells. Then conditionally on the survival of parasites $\{\forall t>0, X_t>0\}$, the number of cells whose quantity of parasites is more than $1$ can not remain bounded:
\be
\label{assez}
\P(\limsup_{t\rightarrow \infty} \# V_t^1= \infty)=1.
\ee
Indeed if $\limsup_{t\rightarrow \infty} \# V_t^1< \infty$ and $X_t$ grows exponentially with rate $g$, then for every $I>0$, we can find some cell $C_I$ which contains more than $I$ parasites. Moreover, denoting by $N_I(t)$ the number of cells at time $t$ which are issued from the same given original cell with $I$ parasites and whose quantity of parasites at time $t$ is more than $1$, we have
$$\lim_{I\rightarrow \infty}\sup_{t>0} N_I(t) =\infty.$$
This gives $(\ref{assez})$. Then   the stopping time
$$T_n:= \inf\{t>0 : \ \# V_t^1\geq n\}$$
is finite a.s. Denoting by $V_t(j)$ the set of cells alive at time $t$ when the root of the tree is taken in $j$, we have
$$\frac{\# \{ i \in V_{T_n+t} : X^i_{T_n+t}\geq \exp(\kappa t)\}}{N_{T_n+t}}\geq  \sum_{j \in V_{T_n}} \frac{\# V_t(j)}{N_{T_n+t}}\frac{\# \{ j \in V_{t}(j) : X^i_{t}(j)\geq \exp(\kappa t)\}}{\# V_{t}(j)}$$
Then  letting $t\rightarrow \infty$ in this inequality and noting that $\# V_{T_n}\geq n$ gives
\Bea
&&\P\big(\limsup_{t\rightarrow \infty} \frac{\# \{ i \in V_t : X^i_t\geq \exp(\kappa t)\}}{N_t} = 0 \ \vert \ \forall t\geq 0: X_t>0 \big) \\
&&\qquad \qquad \quad \leq
\P_{1}\big(\limsup_{t\rightarrow \infty} \frac{\# \{ i \in V_t : X^i_t\geq \exp(\kappa t)\}}{N_t} = 0\big)^n
\Eea
Letting $n\rightarrow \infty$ and recalling that the first part of the proof ensures that  $\P_{x_0}(\limsup_{t\rightarrow \infty} \# \{ i \in V_t : X^i_t\geq \exp(\kappa t)\}/N_t = 0)<1$  leads to
$$\P\big(\limsup_{t\rightarrow \infty} \frac{\# \{ i \in V_t : X^i_t\geq \exp(\kappa t)\}}{N_t} = 0  \ \vert \ \forall t\geq 0: X_t>0)=0.$$
This ensures that  ($\ref{evnonext}$) holds.
\end{proof}

\section{Evolution of the infection  with variable division rate}\label{section4}

We now turn to the case of a variable division rate $r(x)$, meaning that the cell division depends on its infection.
 We are still interested in the proportions of cells with a given number of parasites. The consideration of $Z_t(dx)/\E(N_t)$ as in the previous section  is not useful any longer and  does not give simplification.  Thus, we first give an SDE for the evolution of
\begin{equation}\mu_t=Z_t(dx)/N_t.\end{equation}
Using this equation and the asymptotic behavior of the quantity of parasites in a cell line, we give then some asymptotic results and a conjecture for the case where $r$ is monotone.

\subsection{Evolution of the proportions $Z_t(dx)/N_t$}\label{sectionjefflo}

We begin with writing down an SDE for the evolution equation for $\mu_t$ when $t\geq 0$.

\begin{prop}\label{propproportionsrenormNt}
For every $f\in \Co^2_b(\R_+,\R)$,
\begin{align}
\langle \mu_t,f\rangle= & f(x_0) + \int_0^t \int_{\R_+}\left(gxf'(x)+x\sigma^2 f''(x)\right) \mu_s(dx)\,ds+M^{1,f}_t+M^{2,f}_t\nonumber\\
  & + \int_0^t \int_{\R_+} 2r(x)\frac{N_s}{N_s+1}\left[\int_0^1 f(\theta x)K(d\theta)-f(x)\right] \mu_s(dx)\,ds\nonumber\\
  &+  \int_0^t \frac{\langle \mu_s,r\rangle N_s}{N_s+1} \int_{\R_+} \left[\int_{\R_+} f(y)\widehat{K}_s(\mu_s,dy)-f(x)\right] \mu_s(dx)\, ds,\label{equationEmut}
\end{align}
where $\widehat{K}_s(\mu_s,dy)$ is the probability measure characterized by:
$$\forall f\in \Co_b(\R_+,\R_+),\, \int_{\R_+} f(y) \widehat{K}_s(\mu_s,dy)=\int_{\R_+} \frac{r(y)}{\langle \mu_s,r\rangle}f(y)\mu_s(dy),$$
and where $M^{1,f}_t$ and $M^{2,f}_t$ are two martingales with quadratic variation:
\begin{align}
\langle M^{1,f}\rangle_t=  & \int_0^t \frac{1}{N_s}\langle \mu_s(dx),\, 2\sigma^2 xf'^2(x)\rangle ds\nonumber\\
\langle M^{2,f}\rangle_t=  & \int_0^t \frac{N_s}{(N_s+1)^2} \int_{\R_+} r(x)\int_0^1 (f(\theta x)+f((1-\theta)x)-f(x)
-\langle \mu_s,f\rangle)^2 K(d\theta)\,\mu_s(dx)\, ds.\label{crochetsection4}
\end{align}
\end{prop}

\begin{proof}[Proof of Proposition \ref{propproportionsrenormNt}]The number $N_t$ of cells alive at time $t$, solves the following SDE:
\begin{align}
N_t= & 1+ \int_0^t \int_{\R_+\times \mathfrak{I} \times [0,1]}\ind_{i\in V_{s_-}}\ind_{u\leq  r(X^i_{s_-})}Q(ds,du,di,d\theta).\label{equationNt}
\end{align}
From (\ref{martingalegdepop}) and (\ref{equationNt}), we obtain by Itô's formula for processes with jumps (\eg Ikeda Watanabe Th.5.1 p.67 \cite{ikedawatanabe}) that for every $ f\in \Co^2_b(\R_+,\R)$,
\begin{align*}
\frac{\langle Z_t,f\rangle}{N_t}=  f(x_0) + \int_0^t \int_{\R_+} \left(xf'(x)g+x\sigma^2 f''(x)\right) \frac{Z_s(dx)}{N_s}\, ds+M^{1,f}_t+ J^f_t,\label{equationmut}
\end{align*}
where
$$
M^{1,f}_t=\int_0^t \sum_{i\in V_s}\frac{\sqrt{2\sigma^2 X^i_s}f'(X^i_s)}{N_s}dB^i_s
$$ is a continuous square integrable martingale with quadratic variation:
$$
\langle M^{1,f}\rangle_t= \int_0^t \int_{\R_+} 2\sigma^2 x f'(x)^2 \frac{Z_s(dx)}{N_s^2}\,ds=\int_0^t \int_{\R_+} \frac{2\sigma^2 x f'(x)^2}{N_s}\mu_s(dx)\,ds,
$$ and  $J^f_t$ is the jump part:
\begin{align*}
J^f_t= & \int_0^t \int_{\R_+\times \mathfrak{I} \times [0,1]} \ind_{i\in V_{s_-}, \ u\leq r(X^i_{s_-})}\left(\frac{\langle Z_{s-},f\rangle + f(\theta x)+f((1-\theta)x)-f(x)}{N_{s-}+1}\right.
\nonumber \\
& \qquad \qquad \qquad \qquad  \qquad \qquad \qquad \qquad \qquad
-  \left.
\frac{\langle Z_{s-},f\rangle }{N_{s-}}\right)Q(ds,du,di,d\theta).
\end{align*}Let us denote by $J^f_t=V^f_t+M^{2,f}_t$ the semi-martingale decomposition of $J^f_t$. We aim at rewriting (\ref{equationEmut}) to let the infinitesimal generator of a Markov process appear. The finite variation part $V^f_t$ of $J^f_t$ rewrites as:
\begin{equation}
V^f_t= \int_0^t \int_{\R_+} r(x)\left[\frac{\langle Z_s,f\rangle }{N_s+1}+2 \frac{\int_0^1 f(\theta x) K(d\theta)}{N_s+1}-\frac{f(x)}{N_s+1}-\frac{\langle Z_s,f\rangle}{N_s}\right] Z_s(dx)\,ds
\end{equation}by using the symmetry of $K(d\theta)$. As
$$\frac{\langle Z_s,f\rangle }{N_s+1}-\frac{\langle Z_s,f\rangle }{N_s}=-\frac{\langle Z_s,f\rangle}{N_s(N_s+1)}=-\frac{\langle \mu_s,f\rangle}{N_s+1},$$we obtain:
\begin{align}
V^f_t= & \int_0^t \int_{\R_+} r(x)\frac{N_s}{N_s+1}\left[2\int_0^1 f(\theta x)K(d\theta)-f(x)-\langle \mu_s,f\rangle\right] \mu_s(dx)\,ds\nonumber\\
= & \int_0^t \int_{\R_+} 2r(x)\frac{N_s}{N_s+1}\left[\int_0^1 f(\theta x)K(d\theta)-f(x)\right] \mu_s(dx)\,ds+A,\label{etapetauxfois2}\end{align}
where the last term of (\ref{etapetauxfois2}) rewrites
\begin{align}
 A = &
\int_0^t \int_{\R_+} r(x)\frac{N_s}{N_s+1}\big( f(x)-\langle \mu_s,f\rangle\big) \mu_s(dx)\,ds\nonumber\\
=   & \int_0^t \frac{N_s}{N_s+1}\int_{\R_+} \int_{\R_+} r(x)\big(f(x)-f(y)\big)\mu_s(dy) \mu_s(dx)\,ds \nonumber\\
= & \int_0^t \frac{N_s}{N_s+1} \int_{\R_+}  \langle \mu_s,r\rangle \left( \int_{\R_+} \frac{r(x)}{\langle \mu_s,r\rangle} f(x) \mu_s(dx)
-f(y) \right)\mu_s(dy), \nonumber \end{align}
by using the Fubini theorem and the fact that $\mu_s$ is a probability measure.\\
The bracket of the martingale part is:
\begin{align}
&\langle M^{2,f}\rangle_t \nonumber \\
 & \  =\int_0^t \int_{\R_+} r(x)\int_0^1 \left(\frac{\langle Z_s,f\rangle+f(\theta x)+f((1-\theta)x)-f(x)}{N_s+1}-\frac{\langle Z_s,f\rangle}{N_s}\right)^2 K(d\theta) Z_s(dx)ds\nonumber\\
 & \  = \int_0^t \int_{\R_+} r(x)\frac{N_s}{(N_s+1)^2}\int_0^1 \left(f(\theta x)+f((1-\theta)x)-f(x)-\langle \mu_s,f\rangle\right)^2 K(d\theta) \mu_s(dx)ds. \nonumber
\end{align}
This achieves the proof.\end{proof}

\noindent A probabilistic interpretation of the Markov generator in (\ref{equationEmut}) is in progress \cite{bansayegrahamtran} and described further after (\ref{equationapprox}).

\subsection{Moderate infection  for increasing division rates}\label{section62}

We assume here that $r(x)$ is an increasing function of the quantity of parasites. This means that the more the cell is infected, the faster it divides. Low infected cells divide slower and may even stop dividing if $r(0)=0$. That's why a new regime appears here, between recovery and proliferation of the parasites, where a positive fraction of cells is infected but the quantity of parasites inside remains bounded.
We then say that the infection is $\emph{moderated}$. First, we provide two examples where the infection is indeed moderated:
the organism does not recover but the parasites do not proliferate in the cells. Then, we conjecture a criterion so that the proportion of  cells infected by  more than $A$ parasites tends to zero as $A$ goes to infinity. We illustrate this conjecture with simulations in Section \ref{sectionsimu}. Actually 'moderate infection' could be defined in several ways and we consider then the average quantity of parasites per cell.


\subsubsection{Two simple examples}

We introduce two simple examples, which exhibit new behavior. First, we consider the case where
$$\Theta=1/2 \ \ \t{a.s.}, \quad r(x)=0  \ \  \t{if}  \ x< 2 \  \  \ \t{and}  \ \ \ r(x)=\infty \ \ \t{if} \ \  x\geq 2.$$
 As soon as the quantity of parasites in a cell reaches $2$, the cell divides and the  quantity of parasites in each daughter cell is equal to one. The parasites do not proliferate in the cells since the quantity of parasites in each cell is less than $2$. \\
We now fix the growth rate of parasites $g$ such that the probability that the Feller diffusion $(X_t : t\geq 0)$ reaches $0$ before it reaches $2$ is strictly less than $1/2$. Then the number of infected cells follows a supercritical branching process and grows exponentially
 with positive probability. Conditionally on this event, the proportion of infected cells  does not tend to zero since the non-infected do not divide. Thus the organism does not recover. \\
This example shows that we can have both non recovery and non proliferation of parasites in the cells: this corresponds to the 'moderate infection' we define in Section \ref{secconj}.\\

Second, we focus on the linear division rate:
$$r(x)=\alpha x  \quad (x\geq 0),$$
for some constant $\alpha>0$. Then  a new cell appears at rate $\alpha X_t=\sum_{i \in V_t} r(X_t^i)$
where we recall that $X_t$, the total quantity of parasites at time $t$, follows
a Feller diffusion with drift $g>0$. As a consequence, as soon as the parasites do not become extinct, they grow exponentially
 and the number of cells grows with the same exponential rate. Thus the number of cells and the quantity of parasites  grow similarly, which corresponds to non explosion of the average number of parasites per cell (see Section \ref{section:average}).

\subsubsection{Proportion of cells moderately infected}
\label{secconj}
Using an approximation of the evolution of the proportion of cells based on Prop. \ref{propproportionsrenormNt}, we can conjecture the  following criterion for 'moderate infection'. We are interested in the case where the proportion of cells infected by more than $A$ parasites vanishes (uniformly in time) as $A$ tends to infinity. This means that the quantity of parasites in a cell chosen uniformly at time $t$ does not explode as $t\rightarrow \infty$. The criterion still depends on the maximum division rate
$$r^*=\sup\{r(x) : x\geq 0\}$$
and on the sharing of parasites given by the random fraction $\Theta$. It gives an analogue of the recovery criterion given in the previous section for
constant division rate. Again the  bias phenomenon favors lineages with a large number of divisions, which explains the factor $2$ in the criterion. Roughly speaking, if there exists a level of infection $x_0$ beyond which  the cells  divide fast enough, then the parasites can not proliferate in a positive proportion of the cells and the infection is moderated. This level comes from the constant rate case: $g\leq 2r(x_0)\E(\log(1/\Theta)).$
Simulations are provided as an illustration in the Section \ref{sectionsimu}.

 \begin{conjecture}\label{thinfectionmoderee} We assume that $r$ increases. \\
(i) If $g<2r^*\E(\log(1/\Theta))$,
then we have the following almost sure convergence:
$$\lim_{A\rightarrow \infty} \lim_{t\rightarrow +\infty} \frac{\#\{ i \in V_t : X_t^i\geq A\}}{N_t} =0.$$
(ii) If $g>2r^*\E(\log(1/\Theta))$, then for every $A>0$,
$$\{\limsup_{t\rightarrow +\infty}\frac{\#\{i\in V_t : X^i_t\geq A\}}{N_t}>0\}=\{\forall t>0,\, X_t>0\} \quad \t{a.s.}$$
\end{conjecture}
Let us here give some details on the approximation which leads to this conjecture. \\
First, let us prove that almost everywhere on the set $\{\lim_{t\rightarrow +\infty} X_t=+\infty\}$ we have $\lim_{t\rightarrow +\infty}N_t=+\infty$.
Let $A>0$, $B=\{\lim_{t\rightarrow +\infty}X_t=+\infty ; \sup_{t\geq 0} N_t<A\}$ and $x_1\geq 0$ such that $r(x_1)>0$. Conditionally
on the event $B$, there exists a random time $\tau>0$ such that $\forall t\geq \tau, X_t>A .x_1$. Then for every $t\geq \tau$, there exists one cell with more than $x_1$ parasites, and a new cell is created with rate at less $r(x_1)$. Hence, the number of cells is stochastically lower bounded by a Poisson process of rate $r(x_1)$. So $\P(B)=0$ and this proves the result.\\
When $\lim_{t\rightarrow +\infty}N_t=+\infty$, the bracket of the martingale part of (\ref{equationEmut}) converges to zero and the random fraction $N_t/(1+N_t)$ converges to 1. Thus we neglect them and consider the following deterministic approximation. For every $f\,:\,\big((t,x)\mapsto f_t(x)\big)\in \Co^{1,2}_b(\R_+\times \R_+,\R)$,
\begin{align}
\langle \widetilde{\mu}_t,f_t\rangle = & f_0(x_0) + \int_0^t \int_{\R_+}\left(\frac{\partial f_s}{\partial s}(s,x)+ gx\frac{\partial f}{\partial x}(s,x)+x\sigma \frac{\partial^2 f}{\partial x^2}(s,x)\right) \widetilde{\mu}_s(dx)\,ds\nonumber\\
  & +\int_0^t \int_{\R_+} 2r(x)\left[\int_0^1 f(s,\theta x)K(d\theta)-f(s,x)\right] \widetilde{\mu}_s(dx)\,ds\nonumber\\
  & +\int_0^t \langle \widetilde{\mu}_s,r\rangle \int_{\R_+} \left[\int_{\R_+} f(y) \widehat{K}(\widetilde{\mu}_s,dy)-f(s,x)\right] \widetilde{\mu}_s(dx)\, ds.\label{equationapprox}
\end{align}
It describes the law of the following non-linear jump diffusion process $(\zeta_t : t\geq 0)$ which is our auxiliary process in this case.
The process $(\zeta_t:t\geq 0)$ is a Feller diffusion with the following additional multiplicative jumps. At time $t$, conditionally on $\zeta_t=x$, the process jumps to $\Theta x$ with rate $2r(x)$, and with rate $\E(r(\zeta_t))$ to a state $S_t$ defined by
$$\P(S_t\in \text{d} x)=\frac{r(x)\P(\zeta_t \in \d x)}{\E(r(\zeta_t))}.$$
We can neglect the resampling term given by the jump $S_t$ and we get
a Feller diffusion with multiplicative jump with increasing rate $2r(.)$ as studied in Section  \ref{sectiontauxnonconstant}. The extinction criterion of Proposition \ref{linerincre}  for this process gives the conjecture for the criterion for moderate infection.

\subsubsection{Average number of parasites per cell}\label{section:average}

We now state some sufficient conditions under which the behavior of the average number of parasites per cell is known.

\begin{prop}\label{lemmernonconstant} We assume that $r$ increases. \\
(i) If $r$ is  convex  then:
$$\sup_{t\geq 0} \E\big(X_t/N_t\big)<\infty.$$
(ii) If $g>r^*,$ where $r^*$ has been defined in (\ref{defrstar}), then:
$$\lim_{t\rightarrow \infty}\E\big(X_t/N_t\big)=+\infty.$$
\end{prop}
One can note that in the first case $r^*=\infty$.
\begin{proof}First, notice that $X_t/N_t=\langle \mu_t,x\rangle$. Applying (\ref{equationEmut}) to $f(x)=x$ and using that $K$ is symmetric gives:
\begin{align}\langle \mu_{t},x\rangle= & \langle \mu_0,x\rangle+\int_0^{t}\Big[g\langle \mu_s,x\rangle-\frac{N_s}{N_s+1}\langle \mu_s,r\rangle \langle \mu_s,x\rangle\Big]ds + M^{1,f}_{t}+M^{2,f}_{t},\label{etape6}
\end{align}
Using Jensen's inequality and the fact that $r(.)$ and $x\mapsto xr(x)$ are convex:
\begin{align*}
\E(\langle \mu_s,r\rangle \langle \mu_s,x\rangle) \geq & \E(r(\langle \mu_s,x\rangle)\langle \mu_s,x\rangle)\geq r\big(\E(\langle \mu_s,x\rangle)\big)\E(\langle \mu_s,x\rangle).
\end{align*}
Thus, as $N_s/(1+N_s)\geq 1/2$:
\begin{align*}\E\big(\langle \mu_{t},x\rangle\big)\leq &\langle \mu_0,x\rangle+\int_0^{t}\Big(g-\frac{1}{2}r\big(\E(\langle \mu_s,x\rangle)\big)\Big)\E(\langle \mu_s,x\rangle)\,ds.
\end{align*}Since
  $t\mapsto \E\big(\langle \mu_{t},x\rangle\big)$ is upper bounded by the solution of the differential equation $y'=(g-r(y)/2)y$ started at $\langle \mu_0,x\rangle$
  which is bounded since there exists $x_0>0$ such that $\forall y>x_0,\,g-r(y)/2<0$.\\

Under the assumptions of (ii), since $(N_s/(N_s+1))\langle \mu_s,r\rangle \langle \mu_s,x\rangle \leq r^* \langle \mu_s,x\rangle$, we obtain from (\ref{etape6}) for $t=0$:
\begin{align}
\E\big(\langle \mu_{T},x\rangle\big)\geq \langle \mu_0,x\rangle+\int_0^{T}(g-r^*)\E\big(\langle \mu_s,x\rangle\big)ds,
\end{align} which gives the result since $g-r^*>0$.
\end{proof}

\subsection{Some results for decreasing division rate}\label{section_decreasing}

When the division rate decreases, very infected cells tend to become more infected whereas low infected infected tend to divide more and get rid of their parasites. We provide in Proposition \ref{prop5.3} the following criterion: as soon as a healthy cell appears, the organism recovers, else, parasites proliferate in every cell.\\

We are not interested here in constant division rate and we assume that $r$ is decreasing and that:
\begin{equation}
\exists x_1>0,\quad r(x_1)<r(0).\label{condition_decroissance}
\end{equation}
This means that non infected cells divide faster than other infected cells. 

\begin{prop}\label{prop5.3}
Conditionally on the event $\{\exists t>0,  \exists \ i\in V_t :  \ X_t^i=0\}$, the organism recovers a.s.
$$ \lim_{t\rightarrow +\infty}\frac{\# \{ i \in V_t : X^i_t=0\}}{N_t} = 1
\quad \t{a.s.}$$
Conditionally on the complement event $\{\forall t\geq 0,  \forall \ i\in V_t :  \ X_t^i>0\}$, for every $A\geq 0$,
$$ \#\{ i : X_t^i\leq A\} \stackrel{t\rightarrow \infty}{\longrightarrow }0 \quad \t{a.s}.$$
\end{prop}
\begin{proof} First, if there exists $t_0>0$ and $i_0\in \mathfrak{I}$ such that $X_{t_0}^{i_0}=0$, then
$N_t^0:=\# \{ i \in V_t : X^i_t=0\}$ grows exponentially with rate $r(0)$ after time $t_0$ in the sense that
$$0< \lim\inf_{t\rightarrow +\infty}e^{-r(0)t} N_t^0\leq \lim\sup_{t\rightarrow +\infty}e^{-r(0)t} N_t^0<+\infty.$$
Let us recall the notation $N^*_t= \#\{i \in V_t : X^i_t>0\}$ that we used in previous proofs. We now prove that the number $N^*_t$ of infected cells grows geometrically with a lower rate that under (\ref{condition_decroissance}) is at most:
$$r=\sup_{0\leq \lambda\leq t\leq 1} \{r(x_1)\lambda+(t-\lambda)r(0)+\ln(p_{t-\lambda})\}<r(0),$$
where $p_{t}=\P_{x_1}(X_t>0)<1$ for  $t>0$ is the survival probability of a Feller diffusion (without jump) at time $t$ starting from $x_1$.\\
The set of infected cells at time $t$ can be partitioned into subfamilies according to the first time $\lambda$ at which one of their ancestors' infection is less than $x_1$. In lineages where the infection remains more than $x_1$, the division rate is upper bounded by $r(x_1)$.\\
Let us consider a family of cells for which $\lambda\leq t$. Before this time the number of infected cells in this family grows at most with rate $r(x_1)$. After this time, the parasite population issued from this cell becomes extinct at time $t$ with probability at least $1-p_{t-\lambda}$ since it starts with at most $x_1$ parasites at time $\lambda$. Moreover, after time $\lambda$, conditionally on the survival of the parasites, the number of infected cells grows at most with rate $r(0)$. \\
Hence, for all $x_0>0$ and $t\in [0,1]$,
\begin{align}
\E_{x_0}(N_t^*)\leq x_0 \E^\lambda\Big(  e^{r(x_1)\lambda}.p_{t-\lambda}.e^{r(0)(t-\lambda)} \Big) \leq e^r,
\end{align}where $\E^\lambda$ denotes the expectation with respect to the r.v. $\lambda$. This ensures that
$N_t^*/N_t$ goes to $0$ in $L^1$ and then in probability. Following the proof in the case of constant division rate gives the a.s. convergence. \\

Second, since for every $A\geq 0$,
$$\inf_{0\leq x_0\leq A} \P_{x_0}( \exists 0\leq t\leq 1, X_t=0) >0,$$
we can follow (\ref{utptemarkov}) and conditionally on the event
$$\limsup_{t\rightarrow \infty} \#\{ i : X_t^i\leq A\} \geq 1,$$
there exists $t\geq 0$ and $i\in\mathfrak{I}$ such that
$X_t^i=0$ a.s. This ends up the proof.
\end{proof}

The following question arises: when does a non-infected cell appear ? We provide a sufficient conditions so that it happens a.s., which gives a condition for a.s. recovery depending on $r_*$ defined in (\ref{defr_star(inf)}).
\begin{corol}
If $g\leq r_*\E(\log(1/\min(\Theta,1-\Theta))$, then the organism recovers a.s.
\end{corol}
\begin{proof}
We consider the quantity $Y$ of parasites in the following particular cell line: at each division, we choose the less infected daughter cell. If the mother cell is infected by $x$ parasites, then the cell divides with rate $r(x)$ and the quantity of parasites in each daughter cell
is   respectively $\Theta x$ and $(1-\Theta)x$, so that the less infected one has $\min(\Theta,1-\Theta)x$ parasites. Thus the process
$Y$ follows  a Feller diffusion with multiplicative jumps by $\min(\Theta,1-\Theta)$ with rate $r(.)$. By Proposition \ref{extrdecr}, this process becomes extincted a.s. under the assumption $g\leq r_*\E(\log(\min(\Theta,1-\Theta))$. Thus there exist $t>0$ and $i\in\mathfrak{I}$ such that
$X_t^i=0$ and the previous Proposition ensures that the organism recovers a.s.
\end{proof}

\section{Simulations}\label{sectionsimu}

In this section, we illustrate on simulations the criteria that are exposed in this work. For this, we fix $\sigma$, $K(d\theta)$ and $r(x)$ and let the growth rate $g$ of the parasites vary. For each values of this growth rate, we perform simulations and compute a relevant quantity (for instance the probability of extinction of the parasites). \\

\noindent \underline{Example 1}: Let us first consider the case of a constant division rate $r$. We simulate the process $(\xi_t,t\in \R_+)$ that provides in this case an approximation for a lineage chosen uniformly.
\begin{figure}[!ht]
\begin{center}
\begin{tabular}[!ht]{cc}
(a) & \hspace{1.5cm}(b)  \\
\hspace{-0.5cm}
\includegraphics[width=0.25\textwidth,height=0.15\textheight,angle=0,trim= 2cm 4cm 2cm 2cm]{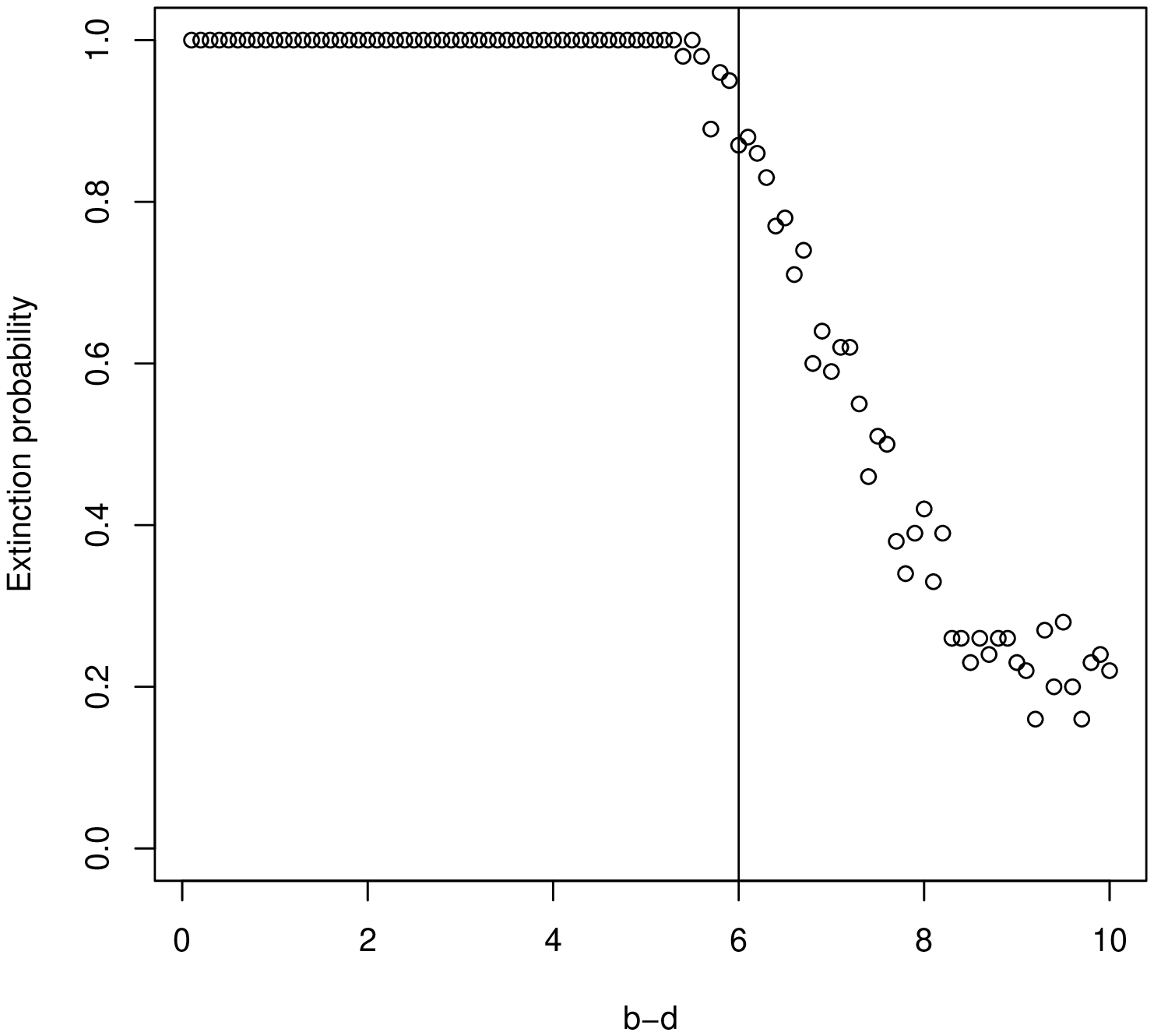} &
\hspace{1.5cm}
\includegraphics[width=0.25\textwidth,height=0.15\textheight,angle=0,trim= 2cm 4cm 2cm 2cm]{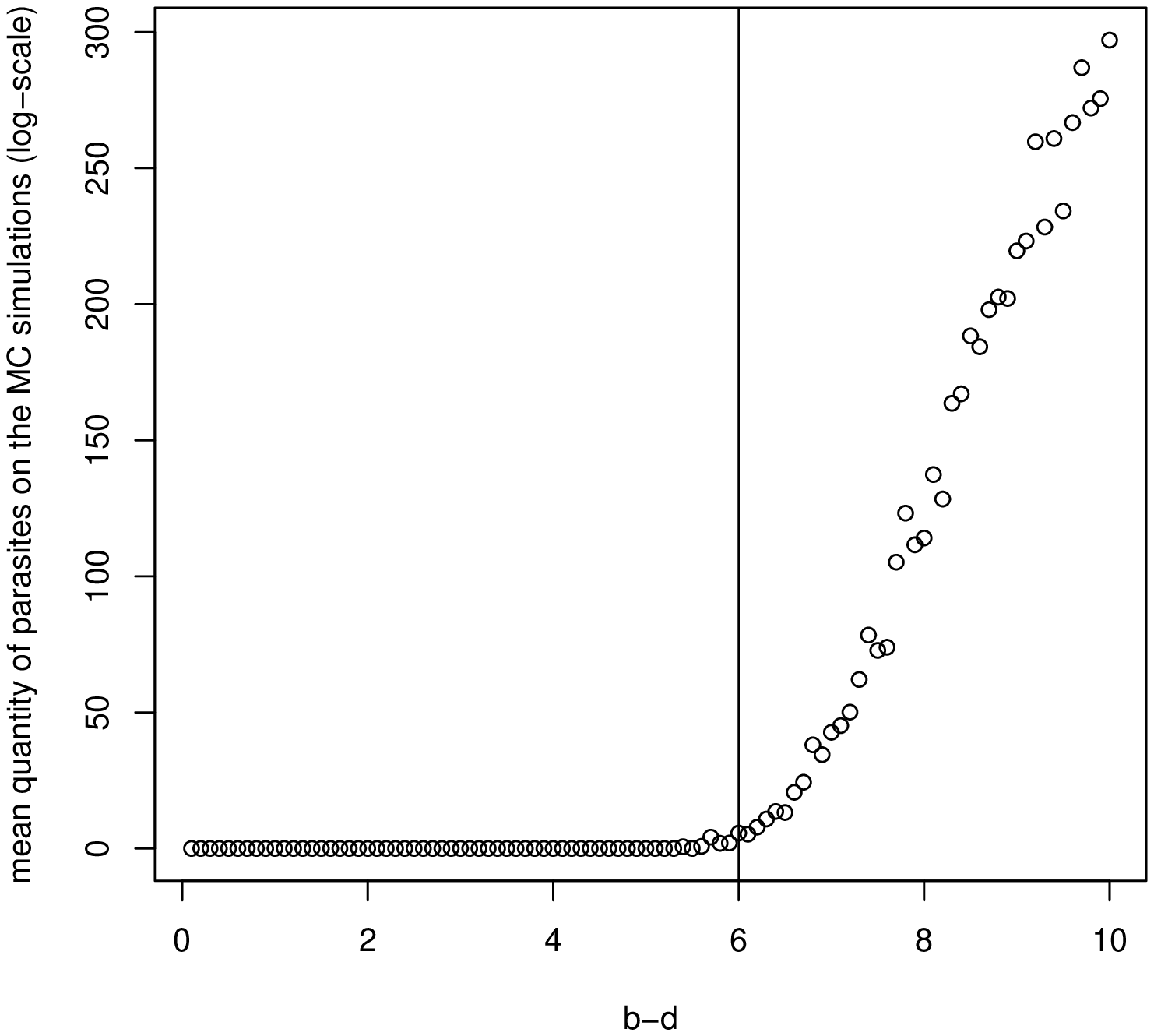}
\end{tabular}
\vspace{0.5cm}
\end{center}
\caption{{\small \textit{Case of a constant division rate: $r=3$, $\sigma=2$. At each division, the random fraction $\Theta$ is drawn uniformly in $[0,1]$. We simulate for each value of $g=b-d$ varying between 0.1 and 10 (abscissa), $N=100$ independent simulations of $(\xi_t,t\in [0,10\,000])$. Their values at $T=10\,000$ provide Monte-Carlo approximations (a) of the extinction probability and (b) of the mean quantity of parasites on a log-scale.
}}}\label{figure_rconstant}
\end{figure}
We consider a random fraction $\Theta$, uniformly distributed on $[0,1]$. Then:
\begin{align*}
\E(\log(1/\Theta))=\int_0^1 \log(1/x)dx=-\int_0^1 \log(x)dx=-[x\log(x)-x]_0^1=1.
\end{align*}In this case, Theorem \ref{threcoveryorganismrconstant} yields that if $g<2r$ the organism recovers a.s. Otherwise, parasites proliferate exponentially with positive probability. We see indeed on the simulations of Figure \ref{figure_rconstant} that recovery happens with probability 1 when $g<2r$ and when $g>2r$, the mean quantity of parasites grows exponentially. \\

\begin{figure}[!ht]
\begin{center}
\begin{tabular}[!ht]{ccc}
(a) & \hspace{1cm}(b)  & \hspace{1cm}(c)\\
\hspace{0cm}
\includegraphics[width=0.25\textwidth,height=0.15\textheight,angle=0,trim= 2cm 4cm 2cm 2cm]{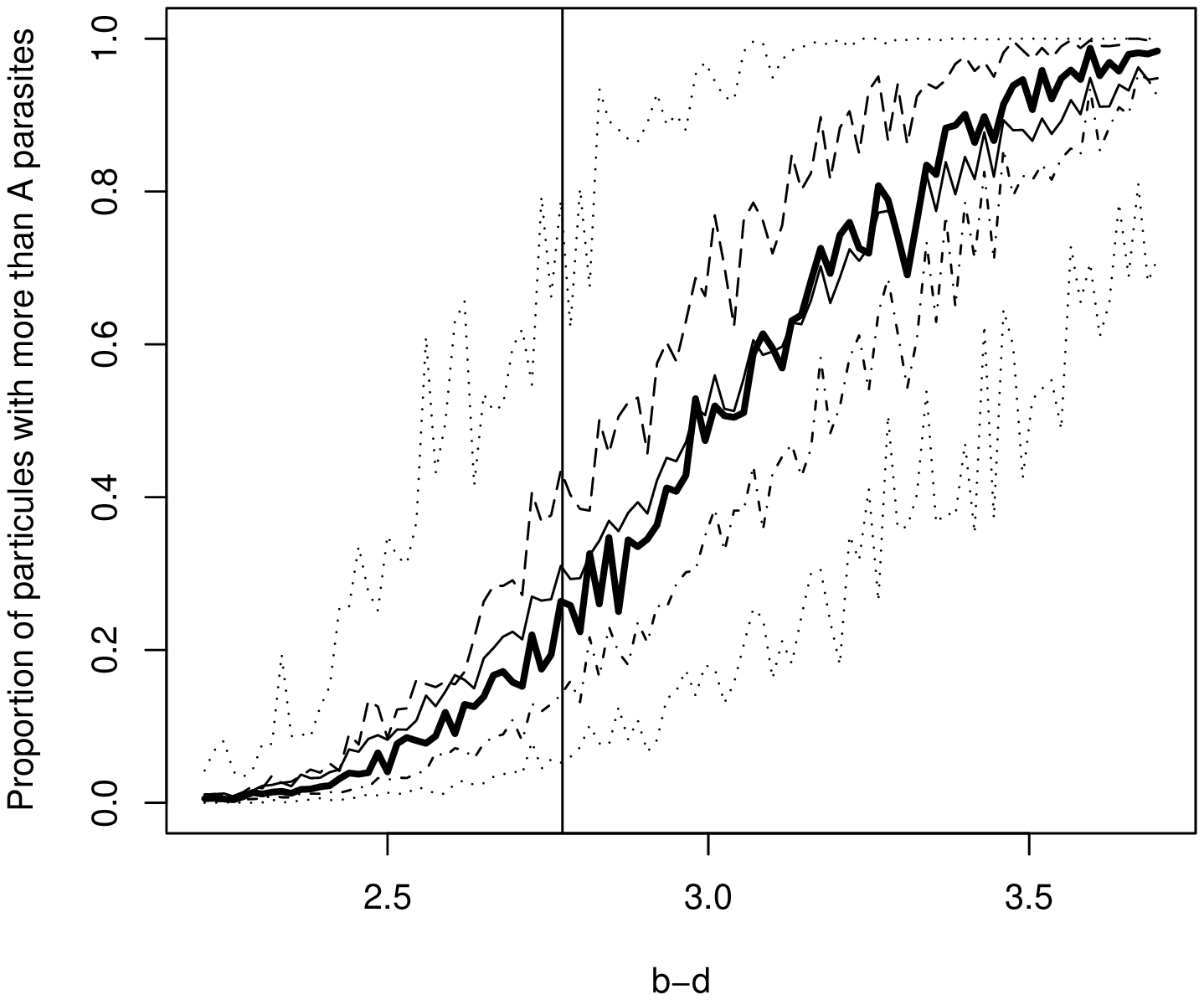} &
\hspace{1cm}
\includegraphics[width=0.25\textwidth,height=0.15\textheight,angle=0,trim= 2cm 4cm 2cm 2cm]{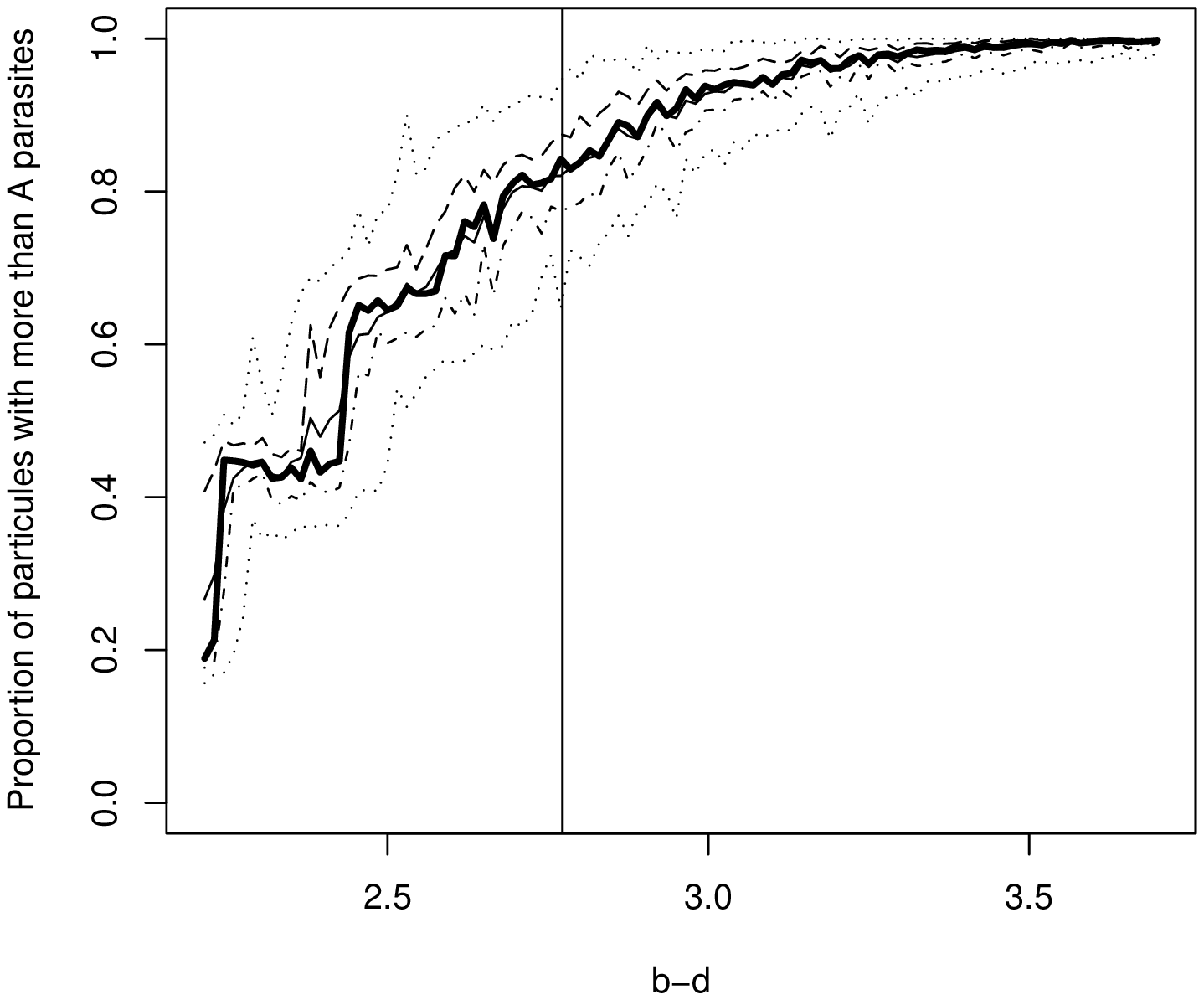} &
\hspace{1cm}
\includegraphics[width=0.25\textwidth,height=0.15\textheight,angle=0,trim= 2cm 4cm 2cm 2cm]{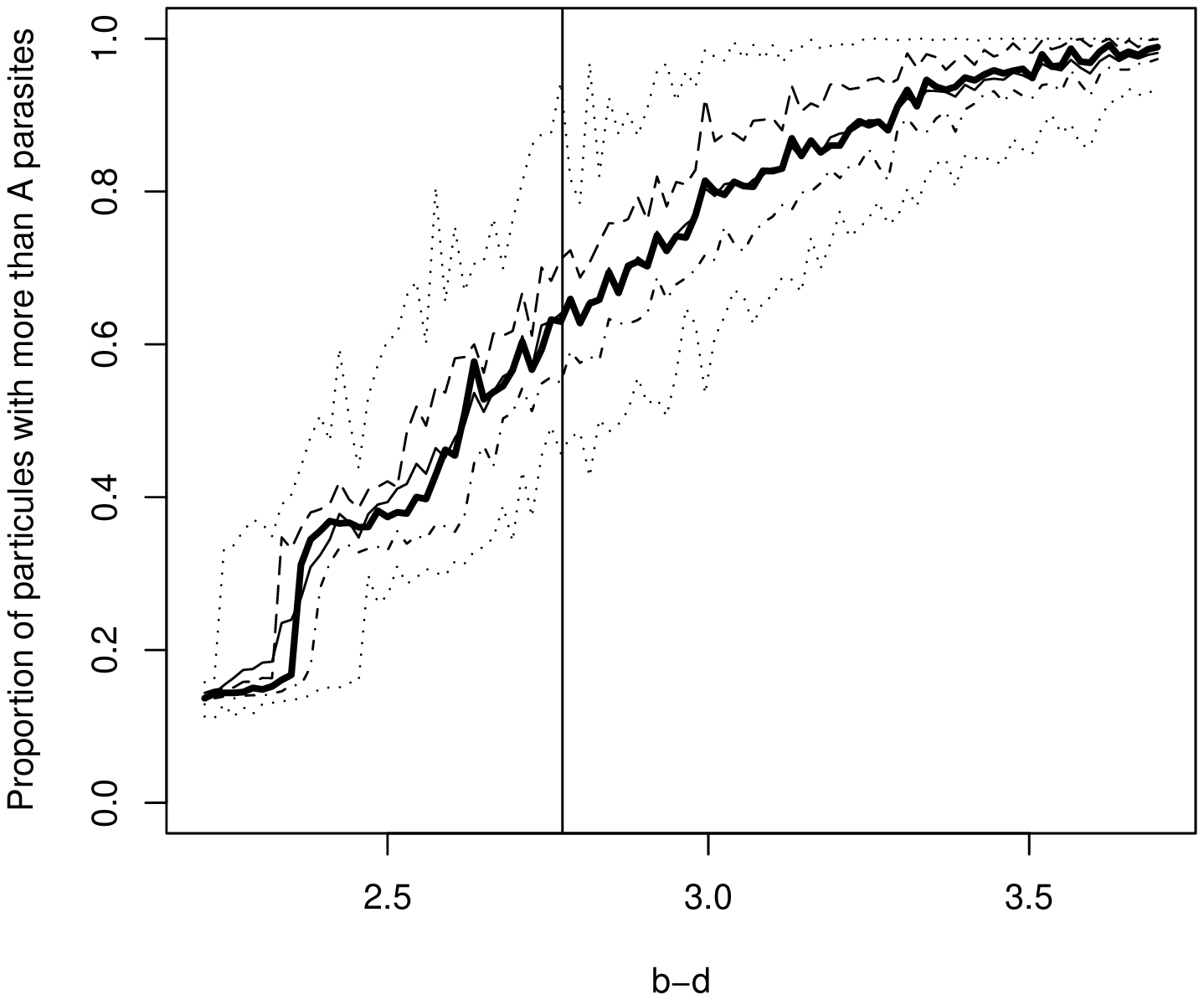}
\end{tabular}
\vspace{0.5cm}
\end{center}
\caption{{\small \textit{Median (Thick plain line), mean (continuous line) and quantiles (2.5\%, 25\%, 75\%, 97.5\%, dotted and dashed) of the distribution of the proportion of cells infected by more than $A=10$ parasites. The bound for the division rate is $r^*=2$ and $K(d\theta)=\delta_{1/2}(d\theta)$. We have simulated the cells up to time $T=700,000$. We let $g=b-d$ vary between 1 and 3.7. For each value of $g$, we simulate 50 branching Feller diffusions   and compute for each simulation the proportion $\#\{i\in V_t : X^i_t \geq A\}/N_t$. This provides an approximation of the distribution of this r.v. for the growth rate $g$. (a)  $r=r^*=2$. (b) $r(x)=r^*(1-exp(-x/10))$. (c) $r(x)=r^*\ind_{[5,+\infty)}(x)$.}}}\label{figure2}
\end{figure}

\noindent \underline{Example 2}: Now, let us illustrate the Conjecture \ref{thinfectionmoderee} that has been made for the case of a variable increasing division rate $r(x)$.
We consider the distribution of the proportion of cells infected by more than $A$ parasites, $\#\{i\in V_t : X^i_t \geq A\}/N_t$, for constant and variable division rates. The real branching diffusions are simulated. Again, we have fixed $\sigma$, $K(d\theta)$ and $r(x)$ and let $g$ vary.\\
We represent in Figure \ref{figure2} (a) the case of a constant division rate. In the simulations of Figures \ref{figure2} (b) and (c), the rate is variable. In these cases, when the infection is low, so is the division rate. The infection hence lasts with a probability that is higher than in the case of Figure \ref{figure2} (a).\\
We can see different behavior depending on whether $g$ is smaller or larger than $2r^*\E(\log(1/\Theta))$, particularly when considering the evolution of the 97\% quantile. \\

\begin{figure}[!ht]
\begin{center}
\begin{tabular}[!ht]{cc}
(a) & \hspace{1.5cm}(b) \\
\hspace{-0.5cm}
\includegraphics[width=0.25\textwidth,height=0.15\textheight,angle=0,trim= 2cm 4cm 2cm 2cm]{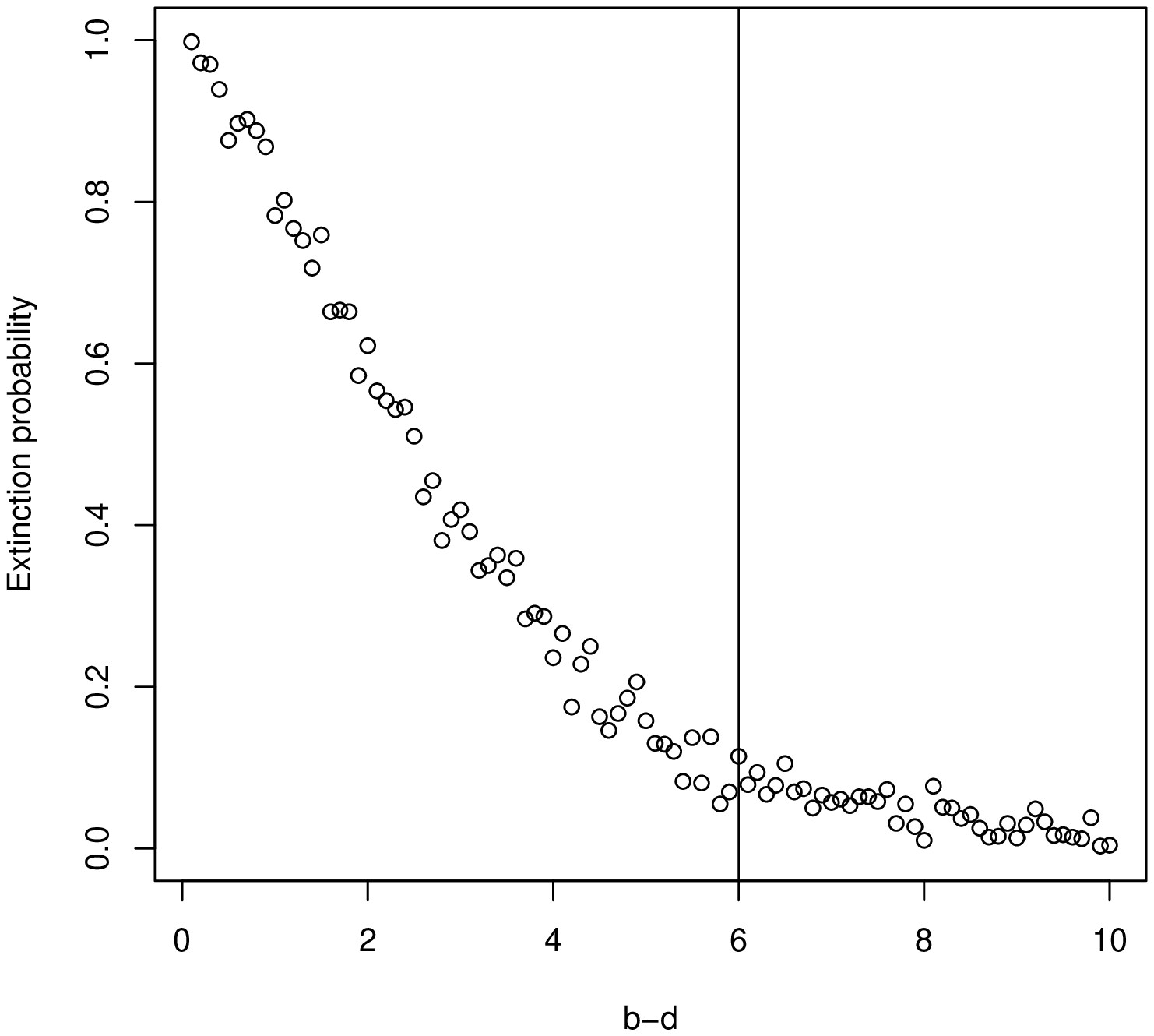} &
\hspace{1.5cm}
\includegraphics[width=0.25\textwidth,height=0.15\textheight,angle=0,trim= 2cm 4cm 2cm 2cm]{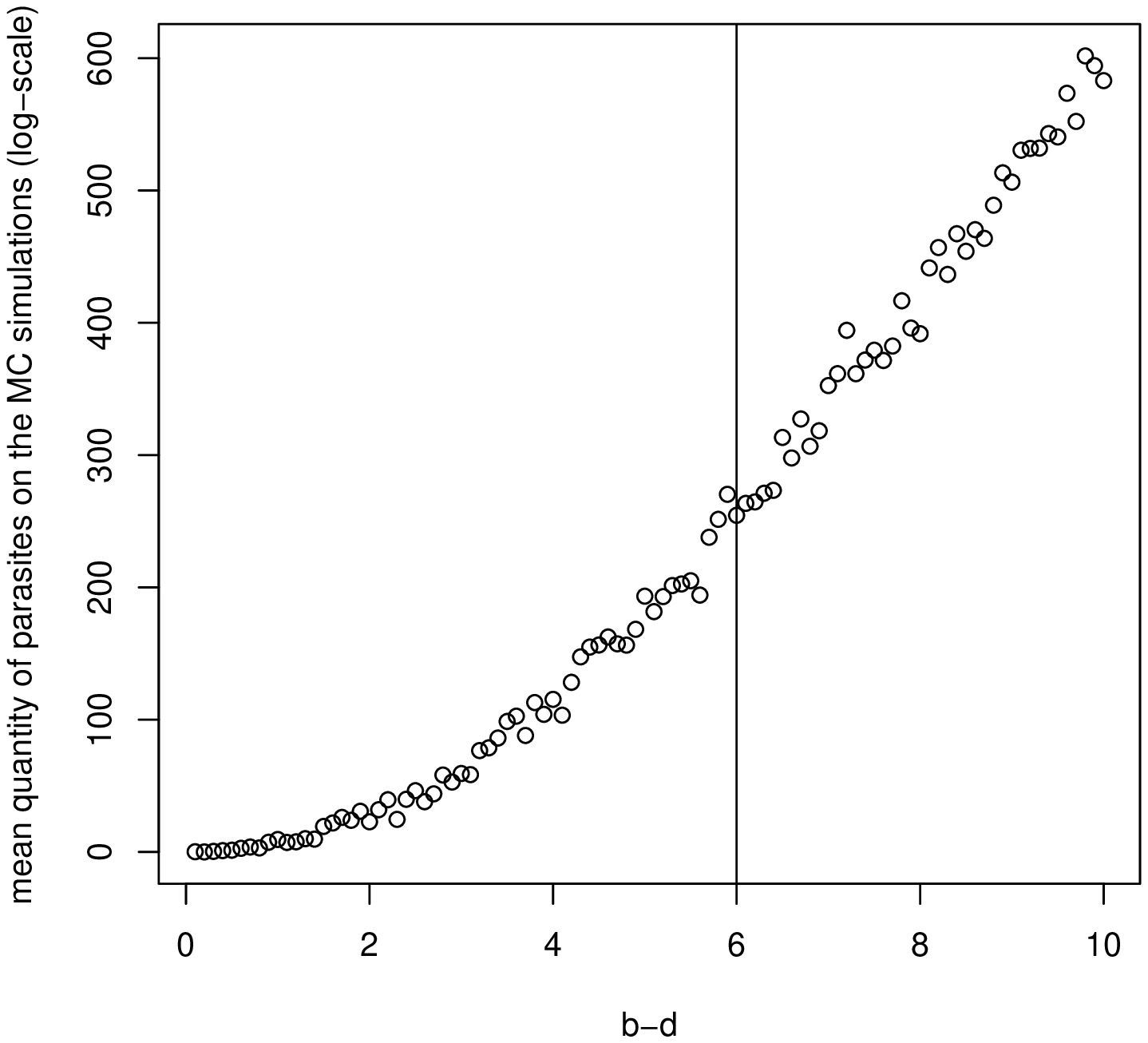}
\end{tabular}
\vspace{0.5cm}
\end{center}
\caption{{\small \textit{Case of a non constant division rate: $r(x)=3\ind_{(2,+\infty)}(x)$, $\sigma=2$. At each division, the random fraction $\Theta$ is drawn uniformly in $[0,1]$. We simulate for each value of $g=b-d$ varying between 0.1 and 10 (abscissa), $N=1000$ interacting particles whose empirical distribution approximates the law of $(\zeta_t,t\in [0,10\,000])$ (see the algorithm in \cite{bansayegrahamtran}). Their values at time $T=10\,000$ provide (a) extinction probability. (b) mean quantity of parasites on a log-scale.
}}}\label{figure_rpalier}
\end{figure}
Although the use of the auxiliary process $(\zeta_t : t\geq 0)$ (Section \ref{secconj}) for proving Conjecture \ref{thinfectionmoderee} has not been fully established yet, the role of the threshold $2r^*\E(\log(1/\Theta))$ can be seen on simulations of $\zeta$ (Figure \ref{figure_rpalier} (a)). Depending on the two regions, we see two regimes for the extinction probability.

\section{Appendix: Proof of Proposition \ref{propconvergence}}\label{appendixmicro}

\begin{proof}We follow Fournier and Méléard \cite{fourniermeleard} and break the proof into several steps. Recall that $Z^n$ is the microscopic process described in Section \ref{section_micro} and that we are under the assumptions of Prop. \ref{propconvergence}. In the sequel, the constants $C$ may change from line to line.\\

\noindent \underline{Preliminaries:} We recall moment estimates that will be useful in the sequel. Recall that $Z^n$ is started with one cell containing $[nx_0]/n$ parasites and that we assumed that $r(x)\leq \bar{r}(1+x^p)$. For all $T\in \R_+$, using computation similar to \cite{fourniermeleard} and stochastic calculus, we prove that for $q\in \N^*$:
\begin{equation}\sup_{n\in \N^*}\sup_{t\in [0,T]}\E(\langle Z^n_s,1+x^q\rangle)<+\infty
\mbox{ and then }\sup_{n\in \N^*}\E(\sup_{t\in [0,T]}\langle Z^n_s,1+x^p\rangle)<+\infty,\label{ptemoment2}\end{equation}
by using Doob's inequality and the first part of (\ref{ptemoment2}) with $q=2p-1$. Let us also notice that:
\begin{align}
&\langle Z^n_t,f\rangle- \langle Z^n_0,f\rangle\label{pbm2}\\
= & \int_0^t \int_{\R_+} \int_{[0,1]}  r(x)
\left[f(\theta x)+f((1-\theta)x)-f(x)\right]K(d\theta)Z^n_s(dx)ds+M^{1,n,f}_t+M^{2,n,f}_t\nonumber\\
+ & \int_0^t \int_{\R_+}Z^n_s(dx)\left[\left(f\big(x+\frac{1}{n}\big)- f(x)\right)(n\sigma^2+b)nx+\left(f\big(x-\frac{1}{n}\big)-f(x)\right)(n\sigma^2+d)nx\right] ds,\nonumber
\end{align} where $M^{1,f}$ and
$M^{2,f}$ are two square integrable martingales starting from 0 and with brackets:
\bea
&& \langle M^{2,n,f}\rangle_t=  \int_0^t\int_{\R_+}\bigg[\left(f\big(x+\frac{1}{n}\big)- f(x)\right)^2(n\sigma^2+b)nx \nonumber\\
&& \t{ }  \qquad \qquad \qquad \qquad  +\left(f\big(x-\frac{1}{n}\big)-f(x)\right)^2(n\sigma^2+d)nx \bigg]Z^n_s(dx)\,ds \nonumber\\
&& \langle M^{1,n,f}\rangle_t=  \int_0^t\int_{\R_+}\int_{[0,1]}r(x)\left[f(\theta x)+f((1-\theta)x)-f(x)\right]^2 K(d\theta)Z^n_s(dx)\,ds, \nonumber\\
&& \langle M^{1,n,f},M^{2,n,f}\rangle_t= 0.\nonumber
\eea
We shall prove that $(Z^n : n\in \N^*)$ is tight in $\mathbb{D}(\R_+,\mathcal{M}_F(\R_+))$, where $\mathcal{M}_F(\R_+)$ is embedded with the topology of weak convergence by using a criterion due to \cite{ethierkurtz}. We will then consider the uniqueness of the limiting values of $(Z^n : n\in \N^*)$ by identifying them as solutions of a certain martingale problem.\\

\noindent \underline{Step 1: Tightness of $(Z^n : n\in \N^*)$} Let us establish tightness on $\mathbb{D}([0,T],\mathcal{M}_F(\R_+))$. For this we apply Theorem 9.1 p 142 of Ethier and Kurtz \cite{ethierkurtz}. We begin to prove that this sequence is tight in $\mathbb{D}([0,T],\mathcal{M}_F(\R_+))$ with the vague topology on $\mathcal{M}_F(\R_+)$. For this, we remark that the set of functions $f\in \Co^2_b(\R_+,\R)$ is a dense subset of $\Co_b(\R_+,\R)$ in the topology of uniform convergence on compact sets and prove that for such functions $f$, the sequences $(\langle Z^n,f\rangle : n\in \N^*)$ are tight in $\mathbb{D}(\R_+,\R)$. In a second time, we prove that the following compact containment condition holds: $\forall T>0,\, \forall \eta>0,\, \exists K_{\eta,T}$ compact subset of $\mathcal{M}_F(\R_+)$,
\begin{equation}
\inf_{n\in \N^*}\P\left(Z^n_t\in K_{\eta,T}, \mbox{ for }t\in [0,T]\right)\geq 1-\eta.\label{compactcontainmentcondition}
\end{equation}

\noindent Let $f\in \Co^2_b(\R_+,\R)$. Using the Taylor-Young formula, there exists $u_1^n(x)$ and $u_2^n(x)$ in $[0,1]$ such that:
\bea
&&\lim_{n\rightarrow +\infty} \left(f\left(x+\frac{1}{n}\right)-f(x)\right)(n\sigma^2+b)nx+\left(f\left(x-\frac{1}{n}\right)-f(x)\right)(n\sigma^2+d)nx \nonumber \\
&=& \lim_{n\rightarrow +\infty}  (b-d)xf'(x)+\frac{\sigma^2 x}{2}\left(f''\left(x+\frac{u_1^n(x)}{n}\right)+f''\left(x-\frac{u_2^n(x)}{n}\right)\right) \nonumber \\
&&\qquad \qquad +  \frac{x}{2n}\left(b\,f''\left(x+\frac{u_1^n(x)}{n}\right)+d\, f''\left(x-\frac{u_2^n(x)}{n}\right)\right) \nonumber \\
& =& (b-d) xf'(x)+\sigma^2 x f''(x).\label{utvarfinie}
\eea Under the assumptions of Prop. \ref{propconvergence}, the finite variation part $V^{n,f}$ of $\langle Z^n,f\rangle$ satisfies:
\begin{align}
|V^{n,f}_t| \leq & 3\|f\|_\infty\bar{r}T\sup_{s\in [0,T]}\langle Z^n_s,1+x^p\rangle \nonumber \\
 & +\left(\|f'\|_\infty(b-d)+\sigma^2 \|f''\|_\infty+\frac{b+d}{2n}\|f''\|_\infty\right) T\sup_{s\in [0,T]}\langle Z^n_s,x\rangle.\label{majovarfinie}
\end{align}By (\ref{ptemoment2}),
\begin{equation}\sup_{n\in \N^*}\mathbb{E}\left(\sup_{t\in [0,T]}|V^{n,f}|\right)<+\infty.\label{tensionetape1}\end{equation}
\noindent Using that
\begin{multline}
 \left(f\left(x+\frac{1}{n}\right)-f(x)\right)^2(n\sigma^2+b)nx+\left(f\left(x-\frac{1}{n}\right)-f(x)\right)^2(n\sigma^2+d)nx\\
 =
 2\sigma^2 x f^{'2}(x)+\frac{C(n,b,d,\sigma,f) x}{n},\label{utmart}
\end{multline}where $C(n,b,d,\sigma,f)=O(1)$ when $n\rightarrow+\infty$, we obtain in the same manner that:
\begin{equation}
\sup_{n\in \N^*}\mathbb{E}\left(\sup_{t\in [0,T]}|\langle M^{1,n,f}\rangle_t+\langle M^{2,n,f}\rangle_t|\right)<
C \sup_{n\in \N^*}\mathbb{E}\left(\sup_{t\in [0,T]} \langle Z^n_t,1+x^p\rangle\right)<+\infty,\label{tensionetape2}
\end{equation}by (\ref{ptemoment2}).
\noindent Let $\delta>0$ and let $((S_n,T_n) : n\in \N^*)$ be a sequence of couples of stopping times such that $S_n\leq T_n\leq T$ and $T_n\leq S_n+\delta$. In the same way that we proved (\ref{majovarfinie}), we can show that:
\begin{align}
\mathbb{E}\left(|V^{n,f}_{T_n}-V^{n,f}_{S_n}|\right)\leq C(b,d,\sigma,\bar{r},f)\delta\, \sup_{n\in \N^*}\mathbb{E}\left(\sup_{t\in [0,T]}\langle Z^n_t,1+x^p\rangle\right)\leq C(b,d,\sigma,\bar{r},f,T)\delta,
\end{align}by (\ref{ptemoment2}). The upper bound can be as small as we wish with a proper choice for $\delta$.
Similarly:
\begin{align}
& \mathbb{E}\left(|\langle  M^{1,n,f}\rangle_{T_n}-\langle M^{1,n,f}\rangle_{S_n}+\langle M^{2,n,f}\rangle_{T_n}-\langle M^{2,n,f}\rangle_{S_n}|\right)\nonumber \\
& \quad \qquad \leq C(b,d,\sigma,\bar{r},f,T)\delta\, \sup_{n\in \N^*}\mathbb{E}\left(\sup_{t\in [0,T]}\langle Z^n_t,1+x^p\rangle\right)\nonumber\\
&  \quad \qquad \leq C(b,d,\sigma,\bar{r},f,T)\delta.
\end{align}
Then  Aldous-Rebolledo and Roelly's criteria \cite{joffemetivier, roelly} ensure that the sequence $(Z^n)_{n\in \N}$ is  tight in $\mathbb{D}(\R_+,\mathcal{M}_F(\R_+))$ where $\mathcal{M}_F(\R_+)$ is embedded with the vague convergence topology.

\par Let us now prove the compact containment condition (\ref{compactcontainmentcondition}) to obtain the tightness in $\mathbb{D}(\R_+,\mathcal{M}_F(\R_+))$ with the weak convergence topology on $\mathcal{M}_F(\R_+)$. Recall that the sets $\mathcal{M}_{\leq N_0}([0,a_0])$ of measures with mass bounded by $N_0$ and support included in $[0,a_0]$ are compact (see \cite{kallenberg}, Sect. 15). Notice that:
\begin{align*}
\left\{Z_t^n\notin \mathcal{M}_{\leq N_0}([0,a_0]),\,t\in [0,T]\right\}\subset \left\{\exists t\in [0,T],\, N_t^n > N_0\right\}\cup \left\{\exists t\in [0,T],\, X^n_t > a_0\right\}
\end{align*}where $N^n_t$ is the number of cells and where $X^n_t$ is the total quantity of parasites at time $t$. Hence:
\begin{align}
\mathbb{P}\big(Z_t^n\notin \mathcal{M}_{\leq N_0}([0,a_0]),\,t\in [0,T]\big)\leq & \mathbb{P}\big(\exists t\in [0,T],\, N_t^n > N_0\big)+ \mathbb{P}\big(\exists t\in [0,T],\, X^n_t >  a_0\big)\nonumber\\
\leq & \frac{1}{N_0}\mathbb{E}\Big(\sup_{t\in [0,T]} N^n_t\Big)+\frac{1}{a_0}\mathbb{E}\Big(\sup_{t\in [0,T]} X^n_t\Big).
\end{align}Thanks to (\ref{ptemoment2}) for fixed $T$, we obtain (\ref{compactcontainmentcondition}) by choosing $N_0$ and $a_0$ sufficiently large. This concludes the proof of the tightness of $(Z^n : n\in \N^*)$ in $\D([0,T],\mathcal{M}_F(\R_+))$.\\

\noindent \underline{Step 2: Identification of the limit} Let us consider an adherence value $Z$ of the sequence $(Z^n : n\in \N^*)$, and let us denote again by $(Z^n : n\in \N^*)$ the subsequence that converges towards $Z$ in law in $\D([0,T],\mathcal{M}_F(\R_+))$. Let $f\in \Co_b^3(\R_+,\R)$. For $k\in \N^*$, let $0\leq t_1< \cdots t_k<s<t\leq T$ and $\varphi_1,\cdots,\varphi_k\in  \Co_b(\mathcal{M}_F(\R_+),\R)$. For $z\in\mathbb{D}([0,T],\mathcal{M}_F(\R_+))$, we define:
\begin{multline}
\Psi(z)=\varphi_1(z_{s_1})\cdots \varphi_k(z_{s_k})\Big[\langle z_t,f\rangle-\langle z_s,f\rangle-\int_s^t \int_{\R_+} \int_{[0,1]}\big(r(x)\big(f(\theta x)+f((1-\theta)x)-f(x)\big)\\
+ xf'(x)(b-d)+\sigma^2 x f''(x)\big)K(d\theta)z_u(dx)\,du\Big].
\end{multline}
Then $
\left|\mathbb{E}\left(\Psi(Z)\right) \right|
\leq   A+B+C
$ where:
\begin{align*}
A= & \left|\mathbb{E}\left(\Psi(Z)\right)-\mathbb{E}\left(\Psi(Z^n)\right)\right|\\
B= & \left|\mathbb{E}\left(\Psi(Z^n)\right)-
\mathbb{E}\left(\varphi_1(Z^n_{s_1})\cdots \varphi_k(Z^n_{s_k})\left[M^{1,n,f}_t-M^{1,n,f}_s+M^{2,n,f}_t-M^{2,n,f}_s\right]
\right)\right|\\
C= & \left|\mathbb{E}\left(\varphi_1(Z^n_{s_1})\cdots \varphi_k(Z^n_{s_k})\left[M^{1,n,f}_t-M^{1,n,f}_s+M^{2,n,f}_t-M^{2,n,f}_s\right]\right)\right|.
\end{align*}
The map $z\in \mathbb{D}([0,T],\mathcal{M}_F(\R_+))\mapsto \Psi(z)$ is continuous as soon as $(t_1,\cdots t_k,s,t)$ does not intersect a denumerable set of points of $[0,T]$ where $Z$ is not continuous (\eg Billingsley \cite{billingsley} Theorem 15.1 p.124). The convergence in distribution of $Z^n$ to $Z$, together with (\ref{ptemoment2}), implies that $A$ converges to 0 when $k\rightarrow +\infty$. Since $M^{1,n,f}$ and $M^{2,n,f}$ are martingales, $C=0$. From (\ref{utvarfinie}),
\begin{align*}
B\leq   & \left|\mathbb{E}\left(\int_0^t\int_{\R_+}\left\{\frac{\sigma^2}{2} x\left[f''\left(x+\frac{u^n_1(x)}{n}\right)+f''\left(x-\frac{u^n_2(x)}{n}\right)-2f''(x)\right]\right.\right.\right.\\
 & + \left.\left.\left. \frac{x}{2n}\left(bf''\left(x+\frac{u^n_1(x)}{n}\right)+df''\left(x-\frac{u^n_2(x)}{n}\right)\right)\right\}Z^n_s(dx)ds\right)\right|
\end{align*}Since $f$ is of class $\Co^3$, the integrand is upper bounded by $Cx/n$. Using (\ref{ptemoment2}), $B\leq C'/n$. This proves that $\mathbb{E}(\Psi(Z))=0$ and hence $W^f$ is a martingale.
The computation of its bracket is standard and obtained by taking the limit in $ \langle Z^n_t,f\rangle^2-\langle Z^n_0,f\rangle^2$ on the one hand, and by using Itô's formula with (\ref{crochetmartingalegdepop}) on the other hand.\\

\noindent \underline{Step 3: Conclusion}
In the Step 2, we have identified the adherence values of the sequence of processes $(Z^n : n\in \N^*)$ as the solutions $Z$ of the martingale problem associated with the following generator $\mathcal{A}$. Following Dawson \cite{dawson}, Section 6.1, we choose for the domain $\mathcal{D}(\mathcal{A})$ the set of functions of the form $F_f(Z)=F(\langle Z,f\rangle)$ with $F\in \Co^2_b(\R,\R)$, $f\in \Co_b^2(\R_+,\R)$ and where $Z\in \mathcal{M}_F(\R_+)$. Let:
\begin{align}
\mathcal{A}F_f(Z)= & F'(\langle Z,f\rangle) \int_{\R_+}\left[(b-d)xf'(x)+\sigma^2 x f''(x)\right] Z(dx)\label{generateur}\\
  + &  F''(\langle Z,f\rangle) \int_{\R_+} \sigma^2 x f^{'2}(x) Z(dx)\nonumber\\
+ &  \int_{\R_+} \int_{[0,1]}  r(x)
\left[F\Big(\langle Z,f\rangle+f(\theta x)+f((1-\theta)x)-f(x)\Big)-F_f(Z)\right]K(d\theta)Z(dx).\nonumber
\end{align}The generator $\mathcal{A}$ is linear. Since it is the infinitesimal generator of the process (\ref{martingalegdepop})-(\ref{crochetmartingalegdepop}), $\mathcal{A}$ is closed by Corollary 1.6 p.10 of \cite{ethierkurtz}. This implies that the resolvent set of $\mathcal{A}$ contains $(0,+\infty)$. By Proposition 3.5 p.178 of \cite{ethierkurtz}, since for any initial condition $Z_0\in \mathcal{M}_F(\R_+)$, (\ref{martingalegdepop})-(\ref{crochetmartingalegdepop}) defines a solution of the martingale problem, the generator $\mathcal{A}$ is dissipative. These last facts imply, by Corollary 4.4 p.187 of \cite{ethierkurtz} that two processes of $\D(\R_+,\mathcal{M}_F(\R_+))$ satisfying the martingale problem associated with $\mathcal{A}$ have the same distribution.
\end{proof}

\noindent \textbf{Acknowledgements} The authors are grateful to Martin Ackermann and Grégory Paul for helpful discussions in biology and  Thomas Lepoutre for his remarks. This work has been financed by ANR MANEGE, ANR Viroscopy and Chaire Modélisation Mathématique et Biodiversité.

{\small

\bibliographystyle{alpha}
\bibliography{biblio}
}
\end{document}